\documentclass[useAMS,usenatbib]{mnras}

\usepackage{natbib}
\usepackage{graphicx}
\usepackage{bm}
\usepackage{amssymb}
\usepackage{amsmath}
\usepackage{verbatim}
\usepackage{mathrsfs}
\usepackage{amsfonts}
\usepackage{latexsym}
\usepackage{color}
\usepackage{units}
\usepackage{array}
\usepackage{relsize}
\usepackage{geometry}
\usepackage{hyperref}
\usepackage{array,booktabs,arydshln,xcolor}
\usepackage{multirow}
\usepackage{nicefrac}

\begin{document}
           
\title[Unbiased constraints on ULA mass from dSph's]{Unbiased constraints on ultralight axion mass from dwarf spheroidal galaxies }
\author[Alma X. Gonzalez-Morales et al]{Alma X. Gonz\'alez-Morales$^{1,2}$\thanks{E-mail: alma.gonzalez@fisica.ugto.mx}, David J. E. Marsh$^{3}$, Jorge Pe\~{n}arrubia$^{4}$, 
\newauthor{L. Arturo Ure\~na-L\'opez$^{2}$}\\
 $^{1}$Departamento de F\'isica, DCI, Campus Le\'on, Universidad de Guanajuato, 37150, Le\'on, Guanajuato, M\'exico\\
 $^{2}$Consejo Nacional de Ciencia y Tecnolog\'ia, Av. Insurgentes Sur 1582. \\
 Colonia Cr\'edito Constructor, Del. Benito   Juárez C.P. 03940, M\'exico D.F. M\'exico\\
$^{3}$ Department of Physics, King's College London, Strand, London, WC2R 2LS, UK\\
$^{4}$ Institute for Astronomy, University of Edinburgh, Royal Observatory, Blackford Hill, Edinburgh EH9 3HJ, UK}\date{\today}
\maketitle

\begin{abstract}
It has been suggested that the internal dynamics of dwarf spheroidal galaxies (dSphs) can be used to test whether or not 
ultralight axions with $m_a\sim 10^{-22}\text{ eV}$ are a preferred dark matter candidate. However, comparisons to theoretical predictions tend to be inconclusive for the simple reason that while 
most cosmological models consider only dark matter, one observes only baryons.
Here we use realistic kinematic mock data catalogs of Milky Way dSph's to show that the ``mass-anisotropy degeneracy'' in the Jeans equations leads to biased bounds on the axion mass in galaxies with 
unknown dark matter halo profiles. In galaxies with multiple chemodynamical components this bias can be partly removed by modelling the mass enclosed within each subpopulation. However, analysis of the mock data reveals that the least-biased constraints on the axion mass result from fitting the luminosity-averaged velocity dispersion of the individual chemodynamical components directly.
Applying our analysis to two dSph's with reported stellar subcomponents, 
Fornax and Sculptor, and assuming that the halo profile has not been acted on by baryons, yields core radii $r_{c}>1.5$ kpc and $r_c> 1.2$ kpc respectively, and $m_a<0.4\times 
10^{-22}\text{ eV}$ at 97.5\% confidence.  These bounds are in tension with the number of observed satellites derived from simple (but conservative) estimates of the subhalo mass function in Milky Way-like galaxies. We discuss how baryonic feedback might affect our results, and the impact of such a small axion mass on the growth of structures in the Universe. 
\end{abstract}

\date{Draft version: \today}

\section{Introduction} 
axion dark matter is described by a classical scalar field, and differs from Cold Dark Matter (CDM, which is described by collisionless particles) on scales below the de Broglie wavelength due to the 
presence of gradient energy \citep[see][for a review]{2017PhRvD..95d3541H,2016PhR...643....1M,2014ASSP...38..107S}. For ultralight axions (ULAs) with $m_a/10^{-22}\text{ eV}\equiv m_{22}\approx 1$ this scale is large enough to be of 
relevance for the cusp-core problem in dSphs, as well as alleviating various other small scale issues with CDM~\citep{2000PhRvL..85.1158H,2014MNRAS.437.2652M,2014NatPh..10..496S,Matos:2000ss,Matos:2000ng,Sahni:1999qe,2000PhRvL..85.1158H}.\footnote{Note that a mass scale of order of $10^{-22} \text{eV}$ has for a long time been a recurring result in the studies
 of axion or scalar field models for galaxy halos and small scale structure, see for instance\citep{1990PhRvL..64.1084P,Sin:1992bg,Sahni:1999qe,Arbey:2001qi,Matos:2000ss,2000PhRvL..85.1158H}.}
 On non-linear scales, axion DM forms a class of pseudo-soliton known as an oscillaton, or ``axion star''~\citep{1969PhRv..187.1767R,Seidel:1991zh,UrenaLopez:2001tw,Guzman:2004wj}. The soliton is supported against gravitational collapse by gradient energy, and is expected to form in the centres of ULA halos. On larger scales, since structure formation proceeds just as for 
CDM, ULA halos should resemble the NFW profile \citep{1997ApJ...490..493N}. Indeed, the NFW profile is found from collisionless $N$-body simulations, which are operationally equivalent to the 
axion model on scales above the de Broglie wavelength~\citep[e.g.][]{Widrow&Kaiser1993,Uhlemann:2014npa}. High-resolution cosmological simulations and other numerical experiments~
\citep{2014NatPh..10..496S,2014PhRvL.113z1302S,2016arXiv160800802V,2016PhRvD..94d3513S} reveal just this: ULA/scalar field DM halos comprise a central soliton core transitioning to an 
NFW-like profile at large radii. The size of the core depends on the axion mass and local density, with larger cores occurring for smaller particle masses and lower densities. Standard CDM halos are 
well described by the NFW profile at all radii and display a central cusp. For almost a decade now, it has been suggested in that the cusp-core problem in dwarfs, as well as other ``small-scale crises'' 
\citep{2015PNAS..11212249W}, may be evidence for DM physics beyond
CDM\citep[e.g.][]{2001ApJ...556...93B,2013PhRvD..87k5007T,2014MNRAS.437.2652M,2016MNRAS.460.1399V,2016PDU....12...56B}. It
is not necessary that a DM model solve all of the apparent small-scale
crises at once (a catch-all solution), but proposed solutions to any
given problem must, of course, be consistent with cosmology and
structure formation.

The stellar dynamics of dwarf spheroidal (dSph) galaxies in the Milky Way (MW) can be used to study the distribution of DM in these systems \citep[see e.g.][for a review]{2013pss5.book.1039W}. dSphs 
are DM dominated at all radii, and so the stars can be seen as test particles orbiting in the DM halo. In particular, Fornax and Sculptor galaxies have two distinct stellar sub-populations of different 
metallicty. \citep{2011ApJ...742...20W} (henceforth, WP11) used the virial quantity $\langle \sigma^2_{\rm los}\rangle$ to measure the DM density profile slope, and showed a preference for cores ($
\rho\propto r^0$) over cusps ($\rho\propto r^{-1}$). 
Different particle physics models for DM predict different halo profiles; therefore the dSph measurements can be used to test the consistency of these models, or even to give evidence for one model 
over another \citep[e.g.][]{2007ApJ...657L...1S}. First attempts to use Stellar dynamics of dSphs  to constrain axion and scalar field DM models are discussed in e.g. \citep{2014PhRvD..90d3517D,2015MNRAS.451.2479M,2016arXiv160609030C} 

In this work we address how stellar velocity measurements in dSph's can be used to place \emph{unbiased} constraints to the dark matter particle mass for an axion DM halo model. We investigate this using 
a series of $N$-body mocks for stars as test particles orbiting in static DM halos. We identify the now-familiar $\beta$-degeneracy, which introduces significant bias in the extraction of halo 
parameters using Jeans analysis when the stellar velocity anisotropy, $\beta$, is unknown. We then show how certain parameters can be extracted in an unbiased way using virial (integrated) 
quantities, where dependence on the anisotropy is reduced. 

\begin{figure}
\vspace{-0.2em}\includegraphics[width=\columnwidth]{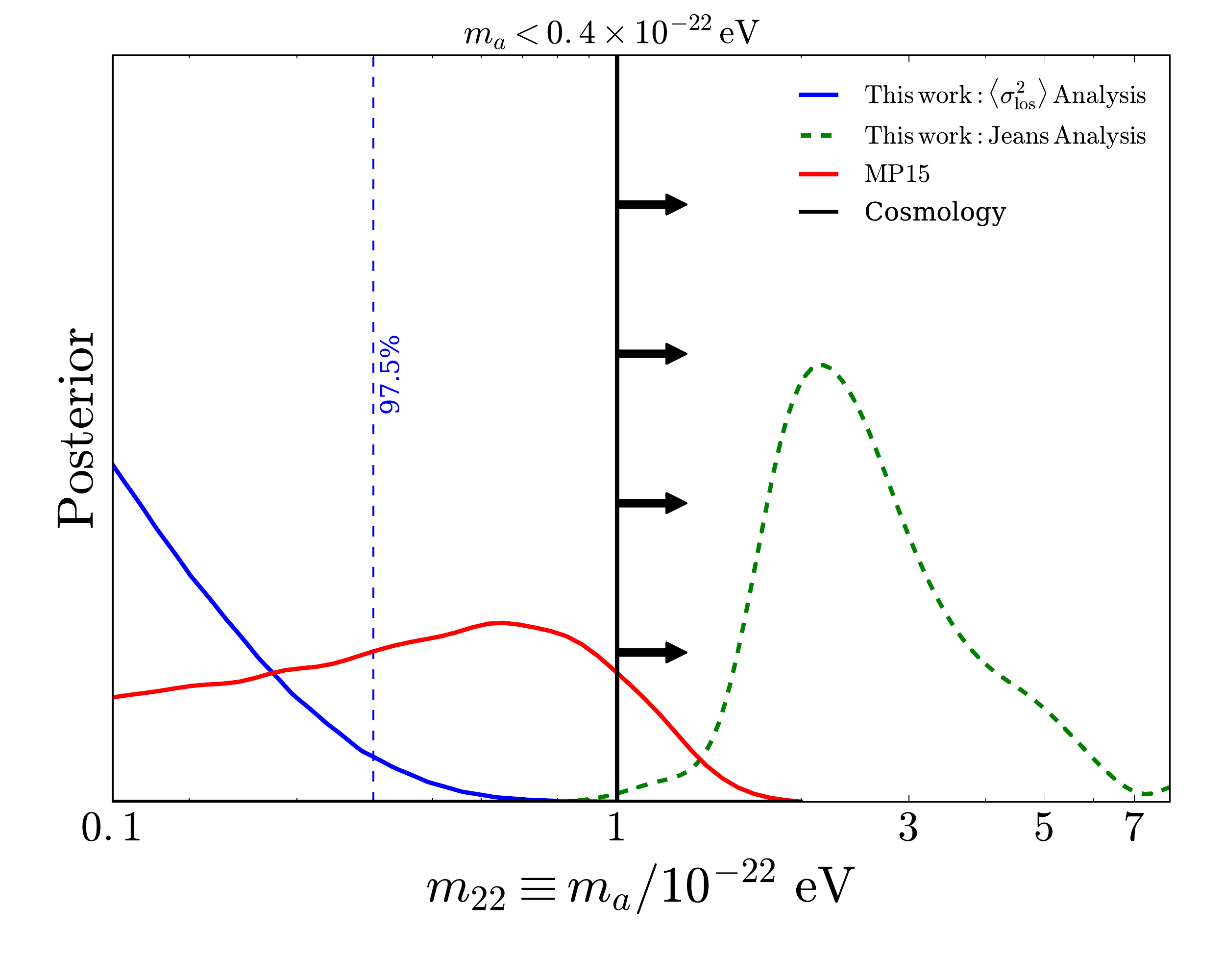}
\vspace{-2.5em}\caption{Marginalized constraint on axion mass from dSph stellar dynamics using three methodologies. Using mock data, we demonstrate that only the $\langle\sigma_{\rm los}
^2\rangle$-fit returns unbiased results. The constraint from this method, $m_a<0.4\times 10^{-22}\text{ eV}$, produces too few subhalos and is inconsistent with a conservative bound of 
$m_a>1\times 10^{-22}\text{ eV}$ from cosmology. Jeans analysis returns the most biased results, due to the anisotropy degeneracy, while the ``virial estimator'' of WP11 used by MP15 (see text) has a slight bias to larger axion masses.}  
\label{fig:mass-lik}
\end{figure}

Fig.~\ref{fig:mass-lik} represents our main findings concerning the axion mass and the MW dSphs. A joint Jeans analysis of the velocity dispersion profile of the eight classical MW dSphs \citep[using 
the data from]{2010ApJ...710..886W} selects a particular axion mass, $m_a=2.44 ^{+1.3}_{-0.6}\times 10^{-22}\text{ eV}$.\footnote{While the present work was in preparation, a similar Jeans analysis 
was performed by \citep{2016arXiv160609030C}, whose results are broadly consistent with ours. We comment on their analysis later.} However, our analysis of mocks leads us to conclude that the 
Jeans analysis has an unknown bias in the recovered axion mass, caused by the $\beta$-degeneracy. Notice that the Jeans analysis is also in some tension with the constraints of 
\citep{2015MNRAS.451.2479M} (hereafter MP15) based on the mass profile slope and virial mass estimator of WP11, which limits  $m_a< 1.1 \times 10^{-22}\text{ eV}$ at a 95\% confidence level 
(C.L.). We revise this upper limit in a new analysis, proved to work extremely well in mock data, finding $m_a< 0.4 \times 10^{-22}\text{ eV}$ at 97.5\% C.L., using $\langle \sigma^2_{\rm los}\rangle$ 
from direct integration of Jeans equation, which we dub the $\langle\sigma_{\rm los}^2\rangle$-fit. 

In the rest of this paper we carefully examine the source of the discrepancy in these bounds on the ULA mass from dSphs, and argue that our revised bound is unbiased. We then discuss possible 
implications from a cosmological perspective. 
This paper is organized as follows. We begin in Section~\ref{sec:cusp_core} by reviewing the status of the cusp-core problem in dSphs. In Section~\ref{sec:model} we describe the ULA halo density 
profile, the model for stellar kinematics, and the set of synthetic observations we use to test our methodology. We perform N-body simulations of stars in the DM potential to  generate mock data. 
 In 
Section \ref{sec:results} we present the results of a Markov Chain Monte Carlo (MCMC) analysis over synthetic observations where we fit the parameters of our model using: ({\it i}) the full velocity 
dispersion profile i.e. Jeans analysis; ({\it ii}) the averaged
velocity dispersion of two stellar populations using a mass-velocity
dispersion  estimator, as it was first proposed in
\citep{2011ApJ...742...20W} and referred here as the slopes method;
and ({\it iii}) a variation of the slopes method, the aforementioned
$\langle\sigma_{\rm los}^2\rangle$-fit, where we propose to use an
unbiased estimator based on a further integration of the Jeans
equation. Then the same three analyses are performed on the real
data. We shall see that $\langle\sigma_{\rm los}^2\rangle$-fit is the
only one provinding unbiased estimates on the axion mass from the kinematics
of dSph's.  Section~\ref{sec:cosmology} discusses the
cosmological (in)consistency of the axion cores, and compares our
constraints to other studies.  We conclude in
Section~\ref{sec:conclusions}. The Appendix presents some additional
details on Jeans analysis, and some discussion of constraints from
cosmological reionization.

\section{Status of the dSphs data, the cusp-core problem, and solutions}
\label{sec:cusp_core}
The dSph satellites of the MW are promising objects to test DM models that differ from CDM on small scales. These old,  pressure-supported systems are the smallest and least luminous known 
galaxies, and there is strong evidence that they are DM-dominated at all radii, with mass-to-light 
ratios as large as \citep{1998ARA&A..36..435M} 
\begin{equation}
M/L_V\sim 10^{1-2}[M/L_V]_{\odot}\,.
\end{equation}

Being the most DM-dominated and metal-poor galaxies in the known Universe \citep{1998ARA&A..36..435M,2012AJ....144....4M}, dSph galaxies play a fundamental role in galaxy formation models 
as well as in investigations of the particle nature of DM. Currently, we lack a clear theoretical understanding of the distribution of DM in these objects. While the density profiles of CDM halos found in 
collision-less $N$-body simulations are well described by a close-to-universal, centrally-divergent (`cuspy') profile \citep[e.g.][]{1997ApJ...490..493N}, alternative DM particle models that allow for 
long-range self-interacting forces \citep[e.g][]{2000PhRvL..84.3760S,2012MNRAS.423.3740V,2016PhRvL.116d1302K} or a small enough DM particle mass \citep[e.g.][]{1979PhRvL..42..407T,2001ApJ...556...93B} 
naturally lead to halo profiles with homogeneous-density `cores'. 

Besides the exotic microscopic properties of DM, a large body of hydro-dynamical simulations suggests that baryons can also reshape the primordial density profiles of dSphs. For example, dense 
baryonic clumps transfer angular momentum to the DM halo as they decay to the inner-most regions of the galaxy through dynamical friction, erasing the central cusp in the process
 \citep{2001ApJ...560..636E,2015MNRAS.446.1820N}. Also, violent periodic fluctuations in the baryonic potential driven by supernova explosions can remove primordial DM cusps \citep{1996MNRAS.283L..72N,
2005MNRAS.356..107R,2012MNRAS.422.1231G,2014MNRAS.437..415D}. Although the amount of supernova energy required to transform the halo profile may become prohibitively large in the 
faintest MW dSphs \citep{2012ApJ...759L..42P}, recently some authors have argued that stochastic star formation in low-mass halos may overcome the energetic limitations, leading to the formation 
of DM cores of size comparable to the stellar half-mass radii of dSphs if star formation proceeds for long enough \citep{2016MNRAS.459.2573R}.  However, other groups using different hydrodynamical codes and feedback recipes do not find DM cores on the mass scale of dSphs at all \citep{2016MNRAS.457.1931S,2016MNRAS.458.1559Z}; the differences seems to be due to a poorer resolution and modelling of the interstellar medium as compared with e.g. \citep{2016MNRAS.459.2573R}. 

The dynamical modelling of dSph galaxies is complicated by the strong degeneracy between the orbital anisotropy of the stellar tracers and the unknown DM distribution in these objects \citep[see]
[for a review]{2013pss5.book.1039W}. These degeneracies arise in the modelling of line-of-sight velocities using the spherical Jeans equations
 \citep[e.g][]{2001ApJ...563L.115K,2007NuPhS.173...15G,2008ApJ...681L..13B,2009MNRAS.394L.102L,2009ApJ...704.1274W,2010ApJ...710..886W,2014MNRAS.441.1584R}, parameterized phase-space distribution functions 
\citep{2002MNRAS.330..792K,2002MNRAS.330..778W,2010MNRAS.408.2364S,2012MNRAS.419..184A,2013MNRAS.429L..89A}, made-to-measure techniques \citep{2010MNRAS.405..301L}, as 
well as orbit-based dynamical models \citep{jardel,2012AAS...21924420J,2013MNRAS.433.3173B,2013A&A...558A..35B}.

WP11 devised a simple method for breaking the mass-anisotropy degeneracy in dSphs with multiple chemo-dynamical populations. For a given dSph, WP11 \citep[see also][]{2012MNRAS.419..184A} use measurements of stellar positions, velocities, and spectral indices to estimate half-light radii and velocity dispersions for as many as two chemo-dynamically independent stellar sub-populations. Several works have shown that the mass estimator \citep{2009ApJ...704.1274W,2010MNRAS.406.1220W}
\begin{eqnarray}\label{eq:mrhalf}
G\,M(<R_{\rm half})=\mu\, R_{\rm half}\,\langle\sigma^2_{\rm los}\rangle\, ,
\label{eq:meanslos_1}
\end{eqnarray}
has a value $\mu$ independent of the (unknown) orbital anisotropy of the stellar population. $R_{\rm half}$ is the projected  half mass radius, which for a Plummer stellar density profile is related to 
the three dimensional one by: $r_{\rm half}=1.305\, R_{\rm half}$,  and the luminosity averaged velocity dispersion is defined as
 \begin{equation}
\langle\sigma_{\rm los}^2\rangle=\frac{\int_{0}^{\infty}{\sigma_{\rm los}^2(R') I(R') R' dR'}}{\int_{0}^{\infty}{I(R') R' dR'}}\,.
\label{eq:meanslos_2}
\end{equation}
Hence, detection of two distinct sub-populations with
different sizes provide mass estimates $M(<R_{\rm half})$ at two different radii in the same mass profile, immediately specifying a slope
\begin{eqnarray}\label{eq:Gamma}
\Gamma\equiv \frac{\Delta \log M}{\Delta \log R_{\rm half}}=1+\frac{\Delta \log \langle \sigma^2_{\rm los}\rangle}{\Delta \log R_{\rm half}}.
\end{eqnarray}
For Fornax and Sculptor, WP11 find slopes of $\Gamma = 2.61^{+0.43}_{-0.37}$ and $\Gamma = 2.95^{+0.51}_{-0.39}$, respectively, which are consistent with cored DM potentials, for which $
\Gamma \le 3$ at all radii, but incompatible with cusped potentials, for which $\Gamma \le 2$.

However, WP11 also showed that the coefficient $\mu$ in Equation~\eqref{eq:mrhalf} is not generally a constant. Tests with mock data reveal that the value of $\mu$ varies depending on (i) the 
spatial segregation of the stellar tracers within the DM halo, and (ii) the DM halo profile itself. In particular, $\mu$ increases as the stellar population is more deeply embedded within the dark matter 
halo, i.e. in the limit $r_{\rm half}/r_{\rm s}\to 0$, and increases more strongly in halos with a shallow density profile. As a result, the values of $\Gamma$ measured by WP11 must be taken as strict 
lower limits, which implies that the exclusion levels of cuspy halo profiles in Fornax and Sculptor are conservative. Moreover, because of the non-constancy of $\mu$, using Eq.~\eqref{eq:mrhalf} to fit 
halo parameters such as the DM particle mass can lead to biased constraints.

\section{Model and synthetic data}
\label{sec:model}

\subsection{The axion halo density profile}

In this work we use the density profile found by~\citep{2014NatPh..10..496S,2014PhRvL.113z1302S,2016PhRvD..94d3513S,2016arXiv160800802V,Mocz:2017wlg} from numerical simulations of structure 
formation with ULAs~(as parameterized in MP15):
\begin{equation}
  \label{eq:density1}  
  \rho(r)= \left\lbrace 
    \begin{array}{cl}
       \displaystyle{\frac{\rho_{\rm sol}}{\left[1+ \left( r/r_{\rm
      sol} \right)^2\right]^8} }\quad & \textrm{for} \quad
      r<r_{\epsilon} \; \\
      \\
       \displaystyle{\frac{\rho_{\rm NFW}}{\left(1+ r/r_{s} \right)^2
      \left( r/r_{\rm s} \right) }} \quad & \textrm{for} \quad r\ge 	r_{\epsilon} \; 
    \end{array}
  \right. \; .
\end{equation}
where $r_{\rm sol}$ is the characteristic radius of the soliton core, and $\rho_{\rm sol}$ the central density. The soliton density and radius are related by the axion mass,
$m_a$~\citep[][MP15]{1969PhRv..187.1767R,2014NatPh..10..496S,2014PhRvL.113z1302S}\footnote{The  soliton profile is an equilibrium configuration that is numerically obtained from the so-called Schrodinger-Poisson 
(SP) system~\citep{1969PhRv..187.1767R,2014NatPh..10..496S,Guzman:2004wj}, and the expression in Eq.~\eqref{eq:density1} is a quite good fitting formula to the numerical solution. Here we are 
following the notation in MP15, but also  see~\citep{2014PhRvL.113z1302S} for an alternative formula. From the original solution of the SP system, the parameters of the soliton profile are explicitly 
given by: $r_{\rm sol} = (0.23 m_a \lambda)^{-1}$, and $\rho_{\rm sol} = m^2_{\rm Pl} m^2_a \lambda^4/4\pi$, where $m_{\rm Pl}$ is the Planck mass, and $\lambda < 10^{-3}$ is a scaling parameter 
(more details can be found in~\citep{1969PhRv..187.1767R,Guzman:2004wj}). Eq.~\eqref{eqn:rsol_scaling} is then obtained from the aforementioned expressions when they are combined to eliminate $\lambda$.}
 \begin{equation}
r_{\rm sol}=\left[\frac{\rho_{\rm sol}}{2.42\times 10^9\; {\rm M}_{\odot} {\rm kpc^{-3}}} \left(\frac{{m_a}}{10^{-22} {\rm eV}}\right)^2\right]^{-0.25}\,{\rm kpc}\, .
\label{eqn:rsol_scaling}
 \end{equation}
 
The parameters corresponding to the external profile (NFW) are $\rho_{\rm nfw}$ and the scale radius $r_{\rm s}$. The radius $r_\epsilon$ is the transition radius from the soliton (inner) to the NFW 
profile (external).
 
We fix the matching radius between the profiles by the density ratio, $\epsilon$, and in turn use this to fix the NFW characteristic density by continuity:
\begin{equation}
\frac{\rho_{\rm sol}}{\left[ 1+ \left( r_{\epsilon}/r_{\rm
          sol} \right)^2 \right]^8}=\epsilon \rho_{\rm sol} =
\frac{\rho_{\rm nfw}}{\left(1+ r_{\epsilon}/r_{s} \right)^2
  (r_\epsilon/r_{\rm s})} \, , \label{eq:continuity}
\end{equation}
We can now rewrite the density profile in the form
\begin{equation}
  \label{density}  
  \rho(r)= \rho_{\rm sol}\left\lbrace 
    \begin{array}{cl}
      \displaystyle{\frac{1}{\left[1+(r/r_{\rm sol})^2\right]^8} }\quad & \textrm{for} \quad
     r<r_{\epsilon}	 \, \\
     \\
       \displaystyle{\frac{\delta_{\rm NFW}}{\left(1+r/r_{\rm
      s}\right)^2 (r/r_{\rm
      s})} } \quad & \textrm{for} \quad r\ge
                                               r_{\epsilon} \, 
    \end{array}
  \right. \, .
\end{equation}
where
\begin{equation}
 r_{\epsilon}=r_{\rm  sol}(\epsilon^{-1/8}-1)^{1/2}\,,
 \label{eq:epsilon}
 \end{equation}
and 
\begin{equation}
 \delta_{\rm NFW}=\epsilon \left[\frac{r_{\epsilon}}{r_{s}}\left(1+ \frac{r_{\epsilon}}{r_{\rm s}}\right)^2\right]\,.
\end{equation}

With this form we can see that the density profile is fully determined once we fix the set of physical parameters:  $m_a [\rm eV]$, $\rho_{\rm sol} [{\rm M}_{\odot} {\rm kpc^{-3}}]$, $\epsilon$ and $r_{\rm s} [{\rm kpc}]$. 
Notice that relation \eqref{eqn:rsol_scaling}  implies that only two of the three parameters that defines de soliton,   $m_a$, $\rho_{\rm sol}$ and $r_{\rm sol}$, are actually independent. In this work we are assuming that there is a universal DM density profile  and dSph's galaxies have not been affected on by barionic feedback. Under such assumptions  we can set $m_a$ to be a universal free parameter in our analysis. We now have some freedom to choose between $\rho_{\rm sol}$ and $r_{\rm sol}$ to be the other free parameter. We decided to use  $\rho_{\rm sol}$ only because the  prior range can be set more intuitively, but as we will show in section \ref{sec:jeans_data} this choice do not affect the results. As we defined the axion mass as a universal parameter, common to all halos, this is essentially a three parameter halo model, with $\epsilon$ the additional parameter over a canonical NFW profile.
There is no definite theoretical prediction on how to set the matching radius $r_{\epsilon}$, though it is expected to be of  order the de Broglie wavelength of the ULA. In the simulations of 
\citep{2014NatPh..10..496S} the transition typically occurs for $\epsilon\sim 10^{-2}$, with a small redshift dependence. In \citep{2014PhRvL.113z1302S,2016PhRvD..94d3513S,2016arXiv160800802V,Mocz:2017wlg},
 soliton mergers are observed to lead to a core-halo mass relationship that in principle determines $\epsilon$ (though details of the results differ somewhat). In practice using a core-halo mass relationship is not efficient for MCMC analysis (as it involves solving an additional integral equation), and so in the present work we take $\epsilon$ as a free 
parameter in each galaxy.\footnote{Note that we choose to work with $\epsilon$, instead of $r_{\epsilon}$, simply because we have a better intuition for the prior: $0< \epsilon <1$ in Eq. \eqref{eq:epsilon}.}
In this model the connection between the galactic dynamics and the properties of DM, i.e. the particle mass, is explicit and once we fix the particle mass by any means it must be the same for all the 
different galaxies in the Universe. On the other hand, the density profile depends on another three parameters that are free to change from galaxy to galaxy. Our expectation is that, observationally at 
least, $\epsilon$ could be correlated with other properties of the cosmological model (e.g. structure formation history),  or with the other free parameters of the halo (central density and scale radius).  
Furthermore, the scaling properties of the soliton suggest $\epsilon$ may be independent of the axion mass.  In the case that all parameters were constrained by the data, one could then use the inferred posteriors to test consistency with the theoretical core-halo mass 
relationship and check the consistency of the inferred dSph density profiles with the formation history in simulations. Since this is not the case, we consider it prudent to simply marginalize over the 
unconstrained degrees of freedom, and focus on constraints to the axion mass.

It is important to state that the purpose of the present work is not to compare the profile in Eq.~\eqref{density} with other halo models in the literature, but rather to use dSph dynamics to investigate 
the parameters of the ULA scenario and test its consistency as an explanation for dSph cores. 

\begin{figure*}
\begin{center}
\includegraphics[width=\columnwidth]{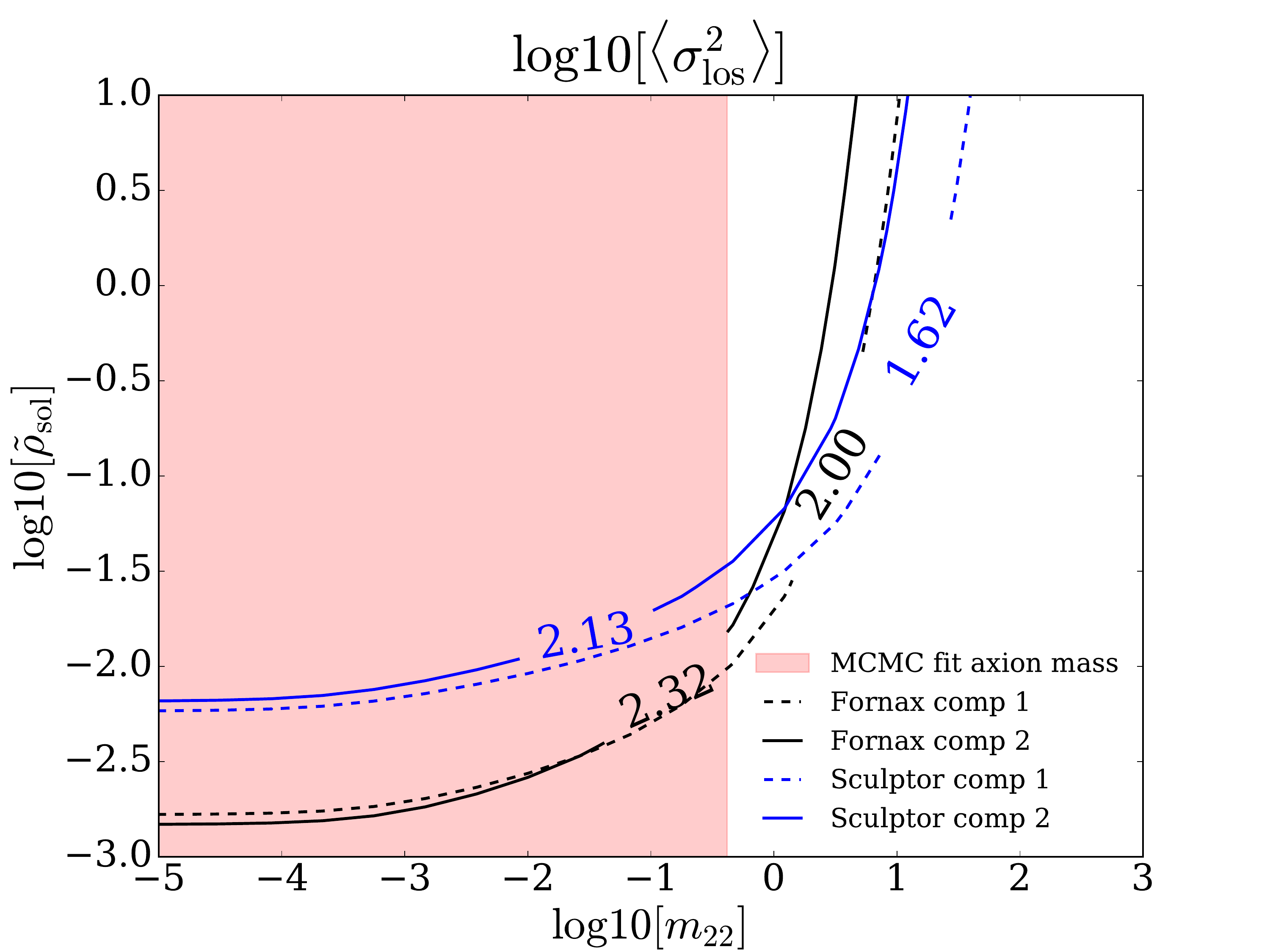}
\includegraphics[width=\columnwidth]{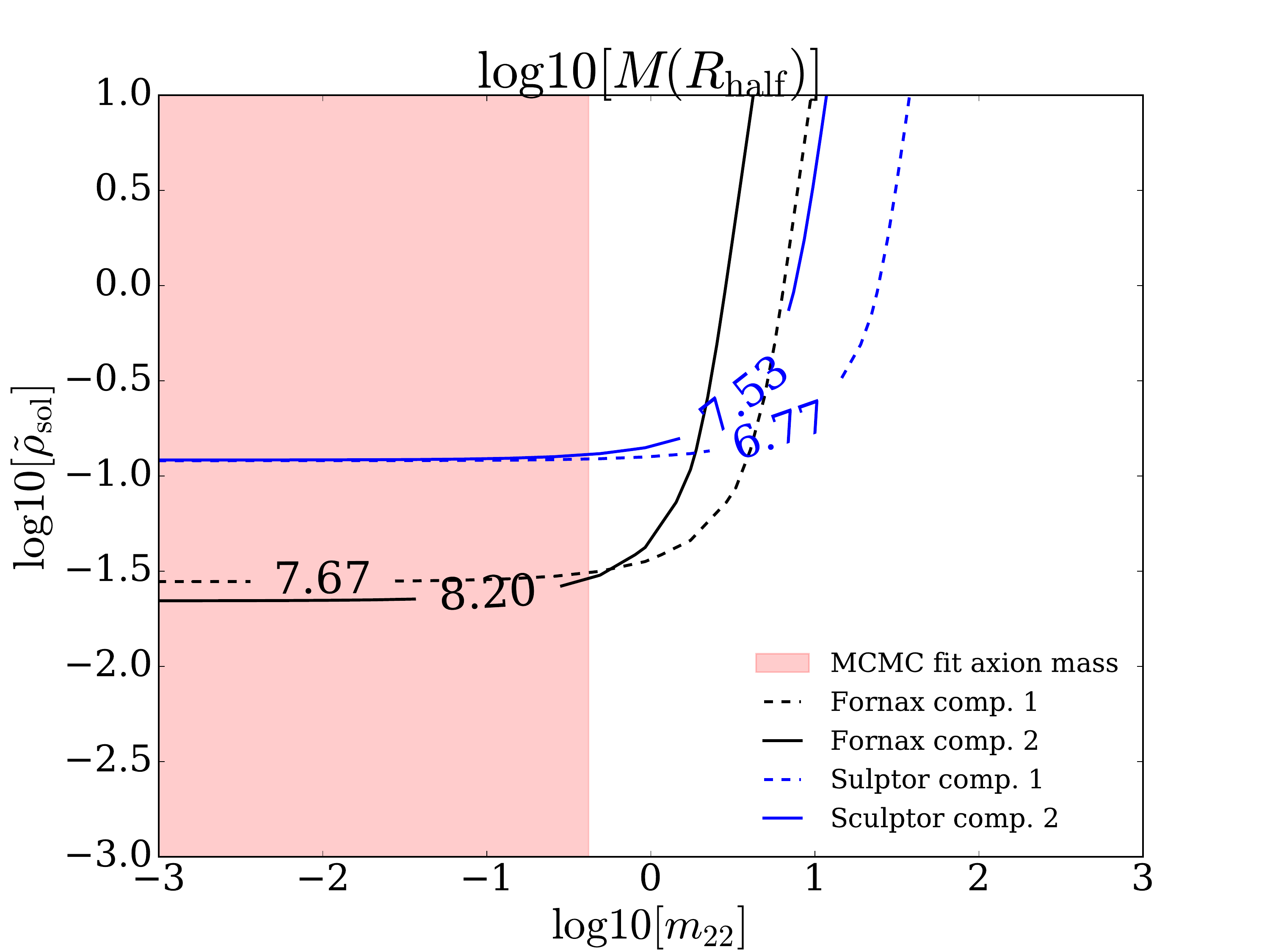}
\caption{ (Left) Mean squared velocity dispersion given by Eq.~ \eqref{eq:meanslos_1},$\langle \sigma_{\rm los}^2\rangle$, for the two observed populations in Fornax and Sculptor galaxies, as a function of the central density and the 
axion mass. (Right) Enclosed mass within the half mass radius, $M(R_{\rm half})$, as a function of the central density and the 
axion mass . Contours correspond to the the median values of $\langle \sigma_{\rm los}^2\rangle$  and $M(R_{\rm half})$  reported in 
\citep{2011ApJ...742...20W}.  Units: axion mass $[10^{-22} {\rm eV}]$; soliton density $[2.42\times10^{9} M_{\odot} {\rm kpc^{-3}}]$; mass $[M_{\odot}]$  and  squared velocity dispersion in  $
[\rm{km^2 s^{-2}}]$.}
\label{fig:slope_WP11}

\end{center}
\end{figure*}
\subsection{Dwarf spheroidal internal dynamics}

\cite{2009ApJ...704.1274W, Walker:2007ju, 2008ApJ...675..201M,2009AJ....137.3100W} reported empirical velocity dispersion profiles for the  eight ``classical'' dSphs of the Milky Way: Carina, 
Draco, Fornax, Leo I, LeoII, Sculptor, Sextans, and Ursa Minor. Our study extends  previous analyses carried out for the generalized Hernquist~\citep{2009ApJ...704.1274W}, Burkert~\citep{Salucci:2011ee} 
and Bose-Einstein condensate~\citep{2014PhRvD..90d3517D} profiles for the DM halo model in Eq.~(\ref{density}). These studies find that all these types of profile provide good fits to the 
data. This is due to the strong degeneracy between the mass density profile and the anisotropy of the velocity dispersion. Here we will apply standard methodologies to generate \emph{mock data}, which reveals 
that without proper knowledge of the true anisotropy, or the true density profile,  the use of Jeans analysis leads to biased constraints on the model parameters.  

Assuming a constant orbital anisotropy, $\beta(r)=\textrm{const}$, the (observed) projection of the velocity dispersion along the line-of-sight, $\sigma_{\textrm{los}}^2 (R)$, relates the mass profile, 
$M(r)$, to the (observed) projected stellar density, $I(R)$ (and the corresponding 3D stellar density $\nu(r)$), through~\citep{Binney}
\begin{equation}
\label{eq:slos_1}
  \sigma_{\textrm{los}}^2(R)=\frac{2G}{I(R)} \int_R^{\infty}{dr' \nu(r') M(r')(r')^{2\beta-2} F(\beta,R,r')} \,, 
\end{equation}
See Appendix~\ref{appendix:jeans} for the specific form of the stellar density, I(R), and the function $ F(\beta,R,r')$ (only dependent on the anisotropy). In section \ref{sec:jeans_mocks}  we use Eq. \eqref{eq:slos_1} to  perform a Monte Carlo analysis to infer the parameters from the line-of-sight velocity dispersion profile, i.e. to perform a standard Jeans analysis. This analysis will show the presence of the $\beta$-degeneracy, leading to significant bias in Jeans analysis.

A second quantity we will examine, also with the help of the mocks, is the luminosity averaged velocity dispersion, $\langle\sigma_{\rm los}^2\rangle$. Being a virial quantity this has the potential to 
yield constraints on the DM density profile that are not affected by
the $\beta$-degeneracy \citep{2011ApJ...742...20W}.

The analysis performed in  WP11 and MP15 uses the empirical relationship given in Eq. \eqref{eq:meanslos_1}, which we call M-estimator, to relate the measured velocity dispersion to the 
enclosed mass. This  relation can be used to set constraints to the model at hand since the density profile, Eq. \eqref{eq:density1}, depends explicitly on the axion mass. However, in WP11 it was also 
shown that this method tends to systematically overestimate the mass of the inner stellar subcomponent to a greater degree than that of the outer stellar subcomponent, and therefore to 
underestimate the slope, with the error introduced depending on the particular DM density profile. 

Fig.~\ref{fig:slope_WP11} shows that the non-constancy of the virial coefficient $\mu$ in Eq. \eqref{eq:mrhalf} affects the axion mass constraints. We draw contours of constant $\langle\sigma_{\rm los}^2\rangle$ (left) and $M(R_{\rm half})$ (right), fixed to the median values reported in WP11 for each population in Fornax and Sculptor, as a function of the axion mass and the central density. The intersection of the solid and dashed black lines corresponds to 
the set of parameters that fit the data for Fornax. Since for Sculptor the blue lines do not intersect, it is Fornax that provides the strongest constraint if we attempt to fit both galaxies simultaneously. In this 
schematic figure we are not considering the confidence interval
reported on the WP11 quantities, and so Fig.~\ref{fig:slope_WP11}
should not be used to estimate constraints on the halo parameters. The
purpose of the figure is to show the difference between fitting the
halo parameters using the averaged velocity dispersion or the enclosed
mass at half light radius.  A comparison of the two panels indicates
that a flat prior on the axion mass can lead to different answers
depending on what quantity is being used to make the fit. Since in
WP11 the observable quantity is $\langle\sigma_{\rm los}^2\rangle$, we
shall  fit to it directly, rather than use the virial estimator for
the enclosed mass. We will show that this choice yields unbiased
constraints on the axion mass.

One final remark is that in
  both panels of Fig.~\ref{fig:slope_WP11} we observe that the axion
  mass cannot be constrained from below, as the two observables
  become sensitive to the central density alone as $m_{22} \to
  0$. That is, the most we can expect is to find an upper bound on the
  axion mass and a well constrained central density, which are just
  consequence of the preference of the WP11 data for large cores. 

In section \ref{sec:sigmalos_mock} we test and compare the fits using the M- estimator and the \emph{averaged velocity dispersion},  Eqs.~\eqref{eq:meanslos_1} and \eqref{eq:meanslos_2}, using mocks of Fornax and 
Sculptor-like galaxies,  each containing two populations of stars with
different half-light radii, for different values of the axion mass and
central density. The analysis using the M-estimator considers
Eq.~\eqref{eq:meanslos_1} and the mass obtained from the density
profile in Eq. \eqref{eq:density1} to do Monte Carlo analysis to infer
the free parameters in the axion model. On the other hand, for the Monte Carlo analysis with the averaged velocity dispersion, also to infer free parameters of the axion model, denoted as $\langle\sigma_{\rm los}^2\rangle$-fit, we use  Eq.~\eqref{eq:slos_1} to compute the squared velocity dispersion profile, $\sigma_{\rm los}^2$, and then find the corresponding luminous averaged 
velocity dispersion as defined in Eq. \eqref{eq:meanslos_2}. The integrals  in Eqs. \eqref{eq:meanslos_2} and \eqref{eq:slos_1} were done using the \emph{quad} routine from the \emph{scipy} library 
\citep{scipy}. Since $\langle\sigma_{\rm los}^2\rangle$ is independent
of $\beta$ we adopt $\beta=0$ for simplicity, and we have corroborated that this works also well in mocks with non-constant orbital anisotropy (see 
Fig.~\ref{mocks} ). 

\subsection{Generation of mock data}
\label{sec:generation-mock-data}
Our model assumes that stars are massless tracers of the DM halo potential, with a spatial distribution chosen so that they describe a Plummer model in equilibrium within the dwarf halo, Eq.~\ref{eq.Plummer}. For simplicity, we assume that the stellar particles have a phase-space distribution function that is spherically symmetric. Regarding anisotropy we construct two set of mocks, one with $\beta=0$ (isotropic), and one in which $\beta$ increases with radius, with as Osipkov-Merrit model. The distribution function is then of the form: 

\begin{equation} 
f(Q)=\frac{1}{8\pi^2} \bigg[ \int_0^Q \frac{d^2\nu_Q}{d\Psi^2}\frac{d\Psi}{\sqrt{Q-\Psi }} +  \frac{1}{\sqrt{Q}}\bigg(\frac{d\nu_Q}{d\Psi}\bigg)_{\Psi=0}\bigg],
\label{fe}
\end{equation}
where $\Psi=-\Phi+\Phi_{\infty}$, $Q=\varepsilon-L^2/2r_a^2$, and $\varepsilon=-E+\Phi_{\infty}$. $\Phi_{\infty}$ is an arbitrary constant that guarantees $\varepsilon \ge 0$ in the radial range of interest, and $\Phi$ is a  solution to the Poisson equation $\nabla^2\,\Phi=4\pi \rho$, where $\rho$ is given by Eq.~\eqref{density}. Notice that
\begin{equation}
\nu_Q(r)=(1+\frac{r^2}{r_a^2}) \nu(r), 
\end{equation}
and that we recover the isotropic case by having very large values of $r_a$. An advantage of our density profile model, Eq.~\eqref{density}, is that there exists 
an analytic solution for the gravitational potential. For a given choice of $\rho(r)$ and $\nu(r)$ we solve Eq.~\eqref{fe} and generate $N_{\star}=10^4$ stellar particles with position and velocity vectors 
$({\bf r}, {\bf v})$ in equilibrium within the DM halo potential.

For the axion model we have two sets of mocks, one with isotropic velocity dispersions, $\beta=0$ and a second one with an Osipkov-Merrit anistropy profile given by:
\begin{equation}
\beta(r)= \frac{r^2}{r_a^2+r^2},
\end{equation}
 
where $r_a$ is  the anisotropy radius, for radius smaller than $r_a$ the velocity dispersion is nearly isotropic, while for larger radius it becomes radially anisotropic. In this second set of mocks we set the anisotropy radius equal to the half mass light radius of the galaxy, i.e $r_a=r_{\rm half}$.
In Tables~\ref{tab:mocks} and \ref{tab:mocksNFW} we specify the parameters that define the mock data for the axion and NFW models respectively. For the axion model we choose a relatively large 
mass of $m_a=2.4\times 10^{-22}\text{ eV}$, consistent with the central value of our joint Jeans analysis performed with the 8 classical dSphs, since we would like to determine under what 
circumstances this model can be reliably recovered from mocks. We  generated mocks for two different half light radii, obtaining the positions and velocities for each population, then this is treated as one single data set, for the analysis of method (i), and issused as the two populations for the analysis of methods (ii) and (iii). \footnote{Mock data is publicly available to use at the GAIA challenge web page \mbox{\url{http://astrowiki.ph.surrey.ac.uk/dokuwiki}}}

\section{Results}
\label{sec:results}

All our results are obtained using Markov Chain Monte Carlo (MCMC) analysis. We optimize the likelihood function (which we define below) to find the maximum likelihood set of parameters, and use 
this as the starting point to explore the parameter space and to estimate the confidence intervals for each parameter. For this task we use the publicly available \textsc{emcee} code~
\citep{Foreman-Mackey}, an affine invariant ensemble sampler, and allow each chain to run up to convergence as defined by the spectral analysis of \citep{2005MNRAS.356..925D}.

We first analyze our mocks, and then move on to the real data. In both cases, we fit $\sigma_{\rm los}$ (Jeans analysis) and $\langle\sigma^2_{\rm los}\rangle$. Details of the Jeans analysis can be 
found in Appendix~\ref{appendix:jeans}; while for the $\langle\sigma^2_{\rm los}\rangle$-fit we use Eq.~\eqref{eq:meanslos_2}, as well as the M-estimator of Eq.~\eqref{eq:meanslos_1}.
\begin{table}
\begin{center}
\begin{tabular}{|ccccccc|}
\hline
Name  &$m_a$&$\rho_{\rm sol}$&$r_{\rm sol}$&$r_{s}$ &$\epsilon$ &  $r_{\rm rhalf}^{\rm comp. 1}$,$r_{\rm rhalf}^{\rm comp. 2}$\\
\hline
\multicolumn{7}{|c|}{{\bf Mock1}}\\
\hline
Fornax &2.44&0.037&1.45& 0.39&0.48&0.549,0.891\\
\hline
Sculptor&2.44&0.057&1.3& 0.57&0.464&0.167,0.302\\
\hline
\multicolumn{7}{|c|}{\bf{Mock2}}\\
\hline
Fornax &0.79&0.03&2.66& 0.39&0.02&0.549,0.891\\
\hline
Sculptor&0.79&0.1&1.99& 0.57&0.06&0.167,0.302\\
\hline
\multicolumn{7}{|c|}{{\bf Mock3}}\\
\hline
Fornax &0.46&0.017&4.08&2&0.01&0.549,0.891\\
\hline
Sculptor&0.46&0.1&2.62&21&0.01&0.167,0.302\\
\hline
\end{tabular}
\end{center}
\caption{Mocks for the axion model. Axion mass and soliton density units are: $10^{-22} {\rm eV}$ and $2.42\times10^{9} M_{\odot} {\rm kpc^{-3}}$, respectively. Soliton and scale radius,($r_{\rm sol}
$,$r_{\rm s}$), are given in kpc. For these parameters we have two sets of mocks, one with $\beta=0$, and one with  $\beta=\nicefrac{r^2}{(r_a^2+r^2)}$ where $r_a=r_{\rm half}$. When used as a single population, the half  light radius is the mean of the two components.}
\label{tab:mocks}
\end{table}

\begin{table}
\begin{center}

\begin{tabular}{|l|c|c|c|}
\hline
Name  &$\rho_{\rm nfw}$&$r_{s}$&$r_{\rm half}$ \\
\hline

 Fornax$_{\rm nfw}$ &0.042&0.49&0.533\\
\hline
 Sculptor$_{\rm nfw}$ &0.0182&0.83&0.235\\
\hline
\end{tabular}

\caption{Mocks for the NFW profile. Density is give in units of $2.42\times 10^9 {\rm M}_{\odot}/{\rm kpc}^3$ and scale radius in kpc. All mocks uses $\beta=0$. }
\label{tab:mocksNFW}

\end{center}
\end{table}

\subsection{Analysis of mock data}
\label{sec:analysis-mock-data}
\begin{figure*}
\includegraphics[width=2\columnwidth]{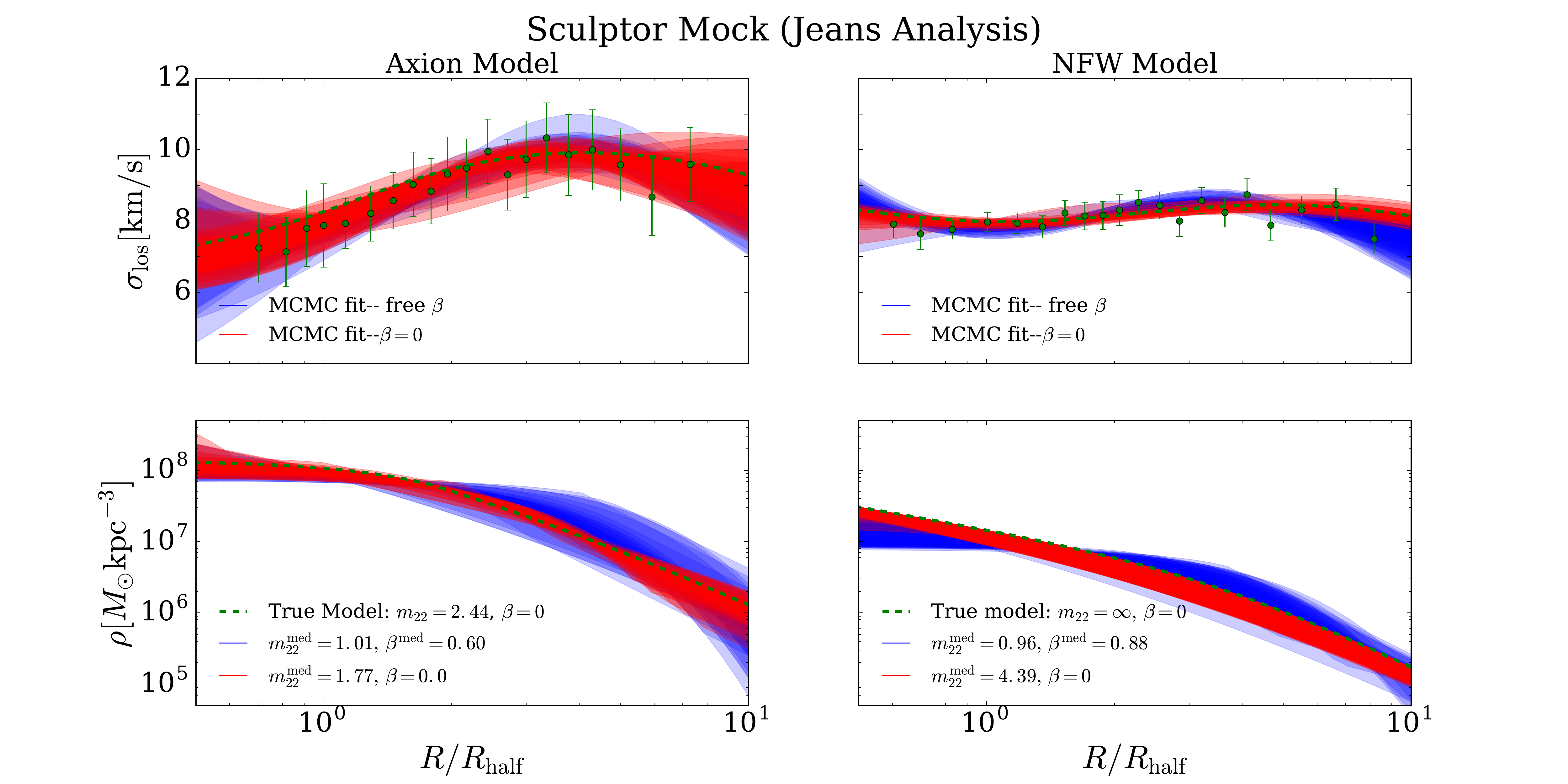}
\caption{The first row shows the velocity dispersion profile for a
  Sculptor-like mock with underlying axion (left), and NFW (right)
  density profiles. For both the axion and NFW mock data we fit the axion
  profile, Eq.~\eqref{eq:density1}, under two separate assumptions:
  (i) the orbital anisotropy, $\beta$, is a free parameter (blue); and
  (ii) the orbital anisotropy is fixed to its true value,
  i.e. $\beta=0$ (red). The second row shows the recovered density profile compared to the true one (green dashed).
Note that for the axion density profile a preferred axion mass value is found but it is smaller than the true value, in particular when $\beta$ is free. For the NFW case with free $\beta$ we find an axion 
mass consistent with a large core, i.e. we find a \emph{false core}, demonstrating the $\beta$-degeneracy. In both cases, only when we set $\beta$ to its known value we do recover a density profile 
close to the true one (green lines). In the NFW mock, the axion model finds the true model as indicated by a power-law density profile $\rho\propto r^{-p}$ with $p>0$, and a large value of the axion 
mass. Axion mass is given in units of $10^{-22} {\rm eV}$.}
\label{fig:scl_}
\end{figure*}

\begin{figure*}
\begin{center}
\includegraphics[width=2\columnwidth]{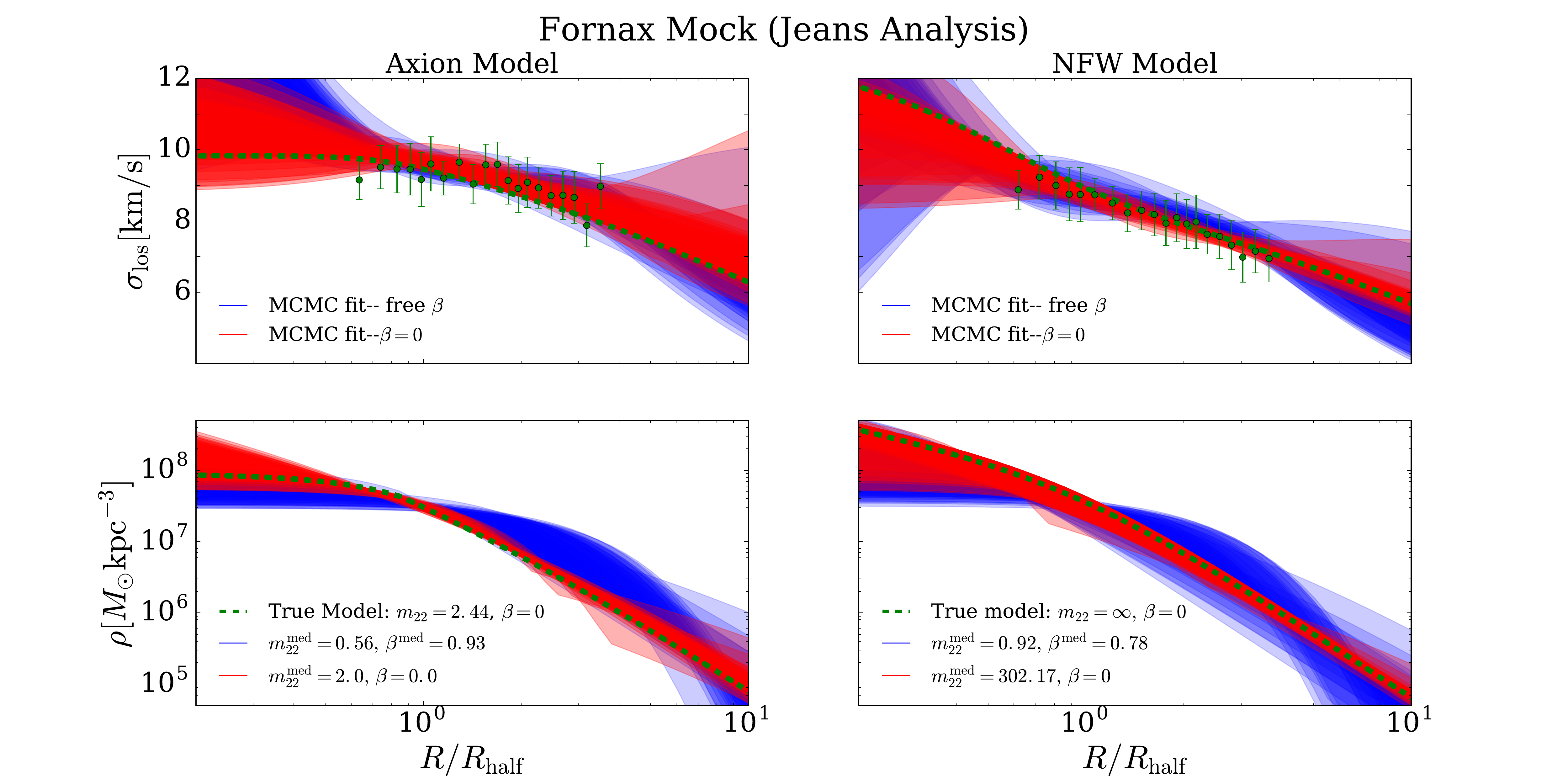} 
 \caption{Same as Fig.\ref{fig:scl_} but for a Fornax-like mock. Again a preferred axion mass smaller than the true value is found when $\beta$ is free (blue). In the case of the NFW mock with free $
\beta$, we find the presence of a large false core. Only when we set $\beta$ to its known value (red) do we recover a density profile and parameters close to the true one (green dashed).}
 \label{fig:for_}
  \end{center}
\end{figure*} 

\begin{figure}
\begin{center}
\includegraphics[width=\columnwidth]{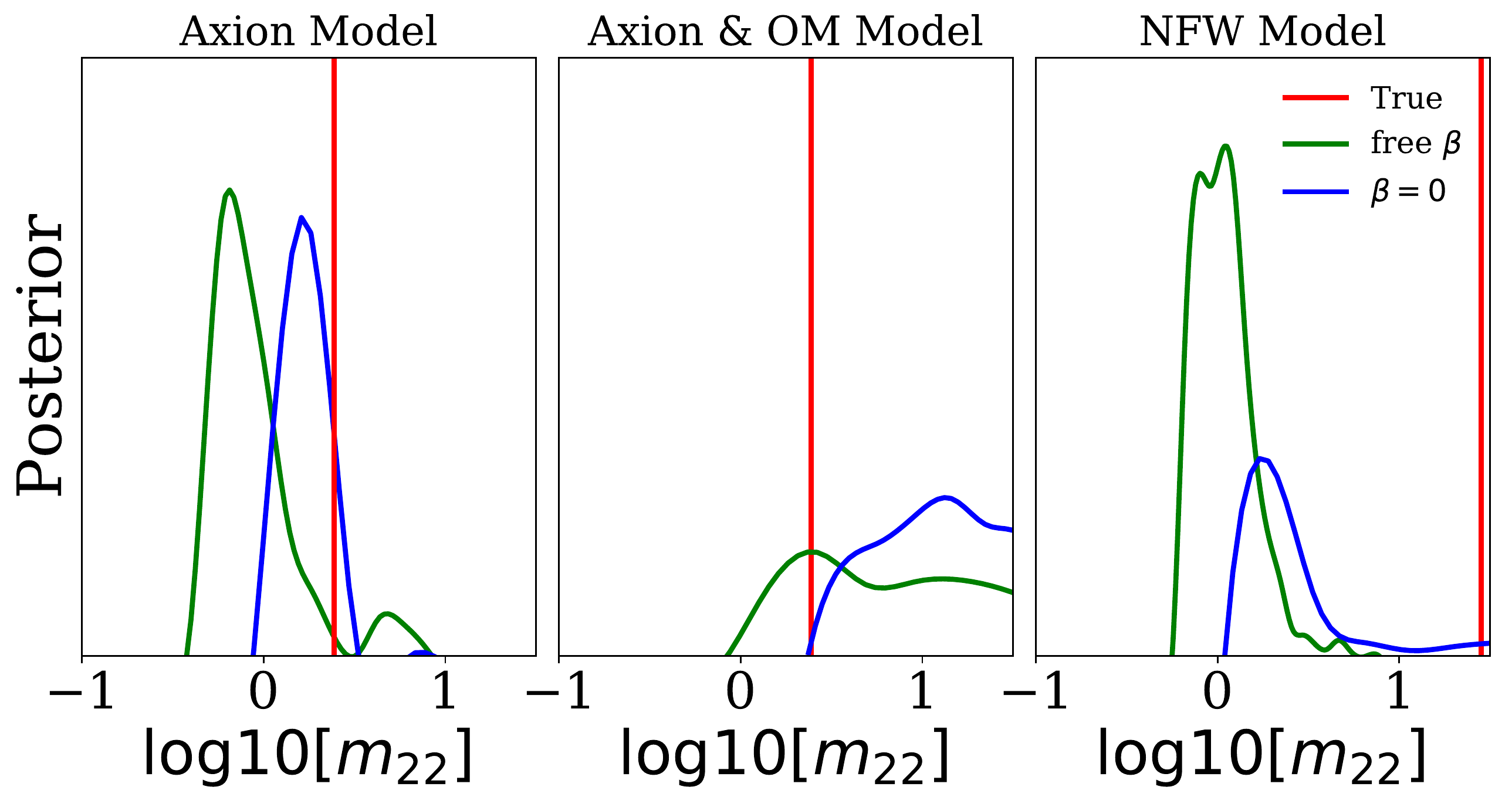} 
 \caption{Posterior distribution for the axion mass from the Jeans analysis in the Fornax mocks for the axion with isotropic velocity dispersion(left), the axion with an Osipkov-Merrit anisotropy velocity dispersion profile (middle),  and NFW density model with isotropic velocity dispersion (right). For isotropic axion (left ) and NFW (right) mocks, the analysis with the anisotropy, $\beta$, set as free parameter tends to recover smaller axion mass (note that the NFW profile is similar to an axion one with large axion mass, Eq. \eqref{eqn:rsol_scaling}, so that the core radius goes to zero). For the isotropic axion (left), only when the analysis is done with the true anisotropy, $\beta=0$, is that we recover an axion mass close to  its true value, but still with a significant bias. For the anisotropic axion (middle) we recover a widely spread posterior, even though the maximum of the posterior coincides with the true value. Finally, the NFW isotropic (right) recovers a false core either with the isotropic or the constant anisotropic model.}
  \end{center}
\end{figure}

\begin{figure}
\begin{center}
\includegraphics[width=\columnwidth]{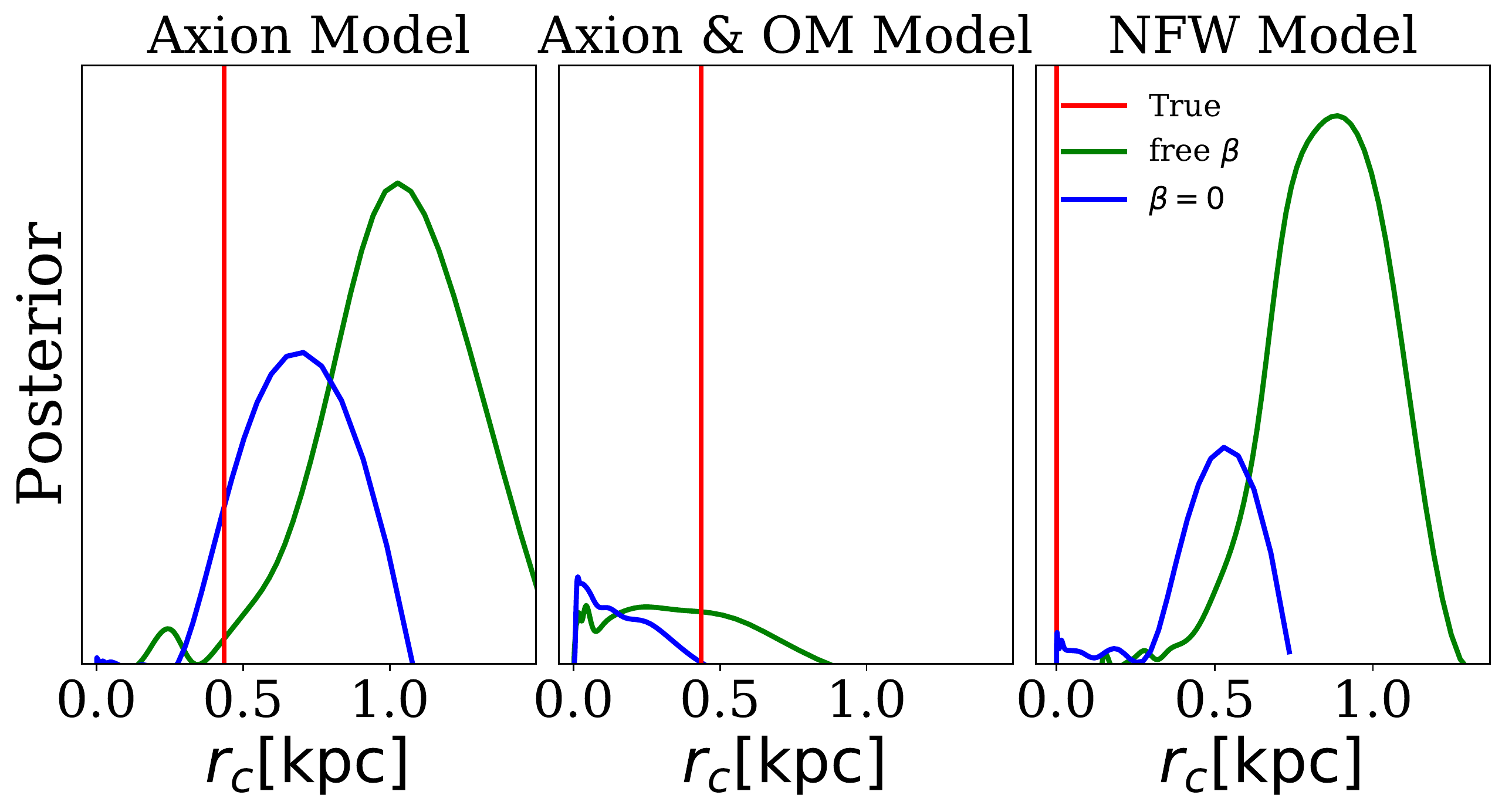} 
 \caption{Same as Fig.~\ref{fig:for_ma} but the for core radius,
   $r_c \approx r_{\rm sol}/ 3$. The analysis with the constant anisotropy, $\beta$, set as free parameter tends to recover a larger core for the isotropic mocks (left and right panels). The analysis recover a core radius closer to its true value when the analysis  is done with $\beta=0$, but still with a significant bias. In the case of  the axion with an Osipkov-Merrit anisotropy profile (middle panel), the analysis either with  $\beta=0$ or  free $\beta=\textrm{const}.$, leads to very spread posterior distributions.  In the case of anisotropic Osipkov-Merrit mock it is interesting to notice how the bias is towards smaller core radii, instead to large core radii, as opossed to the bias in the isotropic mocks.}
 \label{fig:for_rc}
  \end{center}
\end{figure} 

\subsubsection{Jeans Analysis }
 \label{sec:jeans_mocks}
In order to fit the mock observations we have four free parameters per galaxy: three associated with the halo model,  the central density $\rho_{\rm sol}$,  the density matching $\epsilon$ and the 
scale radius $r_{\rm s}$; and one associated with the stellar component, the orbital anisotropy $\beta$. In addition, we also have the mass of the axion $m_a$ which in principle should be a global 
free parameter. However, our first goal is simply to determine how well a Jeans analysis can constrain this parameter in a single galaxy, and so we treat the axion mass as an additional independent 
parameter. We define the likelihood function as:
\begin{equation}
\mathcal{L}(m_a,\bf{\theta}| \sigma)=\prod_{j=1}^{N}{\frac{\exp \left[-\frac{1}{2}\frac{\left(\sigma_{\rm obs}(R_j)-\sigma_{\rm los}(R_j,m_a,{\bf \theta})\right)^2}{{\rm Var}[\sigma_{\rm obs}(R_j)]}\right]}{ \sqrt{ 2 
\pi \, {\rm Var}[\sigma_{\rm obs}(R_j)]}}}\,.
\label{eq:like_jeans}
\end{equation}
where $\theta =(\rho_{\rm sol},\epsilon,r_{\rm s},\beta)$ is the vector of parameters describing the halo model. The index $j$ labels the data bins that runs from $1$ to the total number of bins $N
$. Here $\sigma_{\rm obs}(R_j)$ is the observed line-of-sight velocity dispersion at projected radius $R_j$,  $\sigma_{\rm los}(R_j,m_a,{\bf \theta})$ is given in Eq.~\eqref{eq:slos_1}, ${\rm Var}[\sigma_{\rm obs}(R_j)]$ is the 
square of the error associated with the observed value of the velocity dispersion at $R_j$. See Appendix \ref{appendix:jeans} for more details.

Results are shown in Figs.~\ref{fig:scl_} and~\ref{fig:for_}. In summary: Jeans analysis fails to recover the true density profile (dashed green lines) unless the correct value of the anisotropy is 
adopted (red lines). Using the axion model and allowing the anisotropy to be a free parameter, the Jeans analysis generically infers larger cores than the input ones (blue lines), even recovering a core in the case of an 
NFW input. This can be seen more clearly in Figure~\ref{fig:for_ma} and ~\ref{fig:for_rc},
where we show the posterior distribution for the axion mass and the  core radii respectively (right and leftt panels). Even in
the case where we use the axion model to fit the axion mock data (left
panels in Figs.~\ref{fig:scl_} and~\ref{fig:for_}) and the
correct input $\beta=0$, the Jeans analysis still finds an offsetted
posterior on the axion mass, see Fig.~\ref{fig:for_rc}, although in
this case the true value is within 1-sigma level. In addition, in the middle panels of Figures ~\ref{fig:for_ma} and ~\ref{fig:for_rc} we show the result for the same analyis but performed over the mock with an Osipkov-Merrit  velocity dispersion profile, which is anisotropic. We can see in this case, the analysis with free constant $\beta$ could recover the truth mass since the maximum of the posterior coincides with it, but this will be only at the expense of using a particular  prior. On the other hand, the analysis with $\beta=0$ tends to recover a rather larger axion mass, and consequently a smaller core radius. In the Osipkov-Merrit mocks the resultant  bias is towards larger masses whereas in the isotropic mocks the bias is towards smaller masses. This leave us with the conclusion that in real data a Jeans analysis would lead to biased constraints due to the degeneracy between  the mass profile and the anisotropy profile. 

\subsubsection{Averaged velocity dispersion}
\label{sec:sigmalos_mock}

Now let us analyze the mocks for the axion model, but using the method outlined in WP11. First we will apply the method as it was originally proposed, using  only Eq.~\eqref{eq:meanslos_1}, which we label as M-estimator. We use the data as reported in WP11 where there is also a variance reported. We the use  Eq.~\eqref{eq:meanslos_1}, together with our model,  Eq.~\eqref{density} to make the comparison with data. Second we will adapt the WP11 method to use Eq.~\eqref{eq:meanslos_2} together with Eq.~\eqref{eq:slos_1},  labeled as $\langle\sigma^2_{\rm los}\rangle$-fit. It is important to stress the difference lays in the way we compute the predicted value of $\langle\sigma^2_{\rm los}\rangle$, in the first case we do it through the estimator  given in Eq.~\eqref{eq:meanslos_1}, and in the second one we compute it from a full integration of Jeans equations.
In both cases we do a joint analysis of the two populations in each galaxy. We define the likelihood function as: 

\begin{equation}
\mathcal{L}(m_a,{\bf \theta}| \sigma)=\prod_{gal.}\prod_{j=1}^{2}{\frac{\exp \left[-\frac{1}{2}\frac{\left(\langle \sigma^2_{\rm obs}\rangle_{\rm pop_j}-\langle \sigma^2_{\rm los}(m_a, {\bf \theta}_{gal})\rangle_{\rm pop_j}\right)^2}{{\rm Var}[\langle \sigma^2_{\rm obs}\rangle_{\rm pop_j}]}\right]}{ \sqrt{ 2  \pi \, {\rm Var}\langle \sigma^2_{\rm obs}\rangle_{\rm pop_j}}}}.
\label{eq:like_slope}
\end{equation}
where in this case ${\bf \theta}_{gal}=(\rho_{\rm sol},\epsilon,r_{\rm
  s})$ is the vector of parameters describing the halo model of each
galaxy. Since we are only using the data from two galaxies,  we have 7
free parameters, namely: the axion mass, $m_a$; the central density of
each galaxy, $\rho_{\rm sol}^{\rm For}$, $\rho_{\rm sol}^{\rm Scl}$; and the matching scale and scale radius for each galaxy, $\epsilon^{\rm 
For}$,$\epsilon^{\rm Scl}$,$r_{s}^{\rm For}$, $r_{s}^{\rm Scl}$.  The last four parameters tend to be completely unconstrained (consistent with the analysis of MP15), and therefore we will not include 
them in the triangle plots of posteriors, nor in the discussion.  

\begin{figure*}
\begin{center}
{\includegraphics[width=0.48\textwidth]{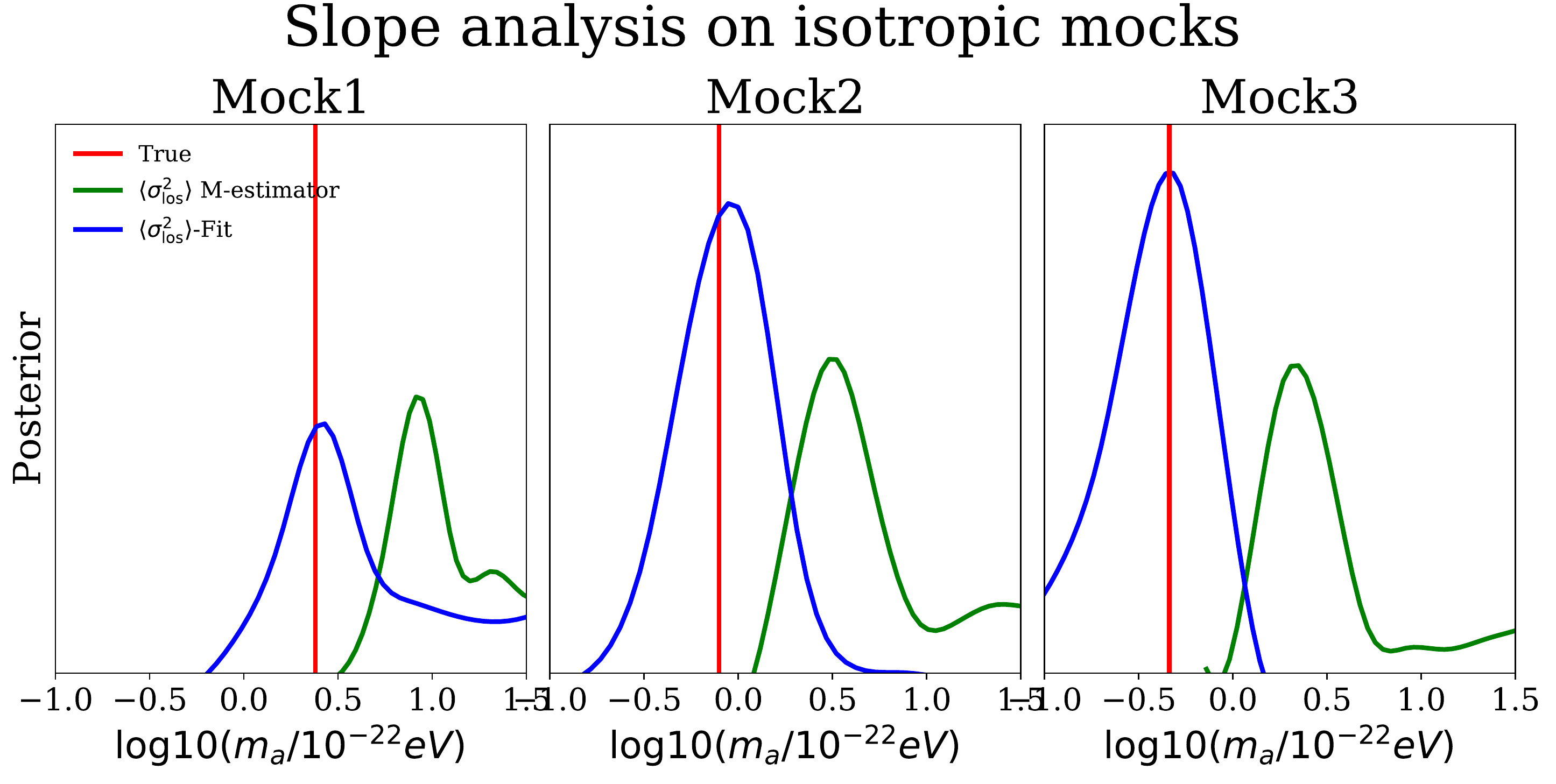} 
  \includegraphics[width=0.49\textwidth]{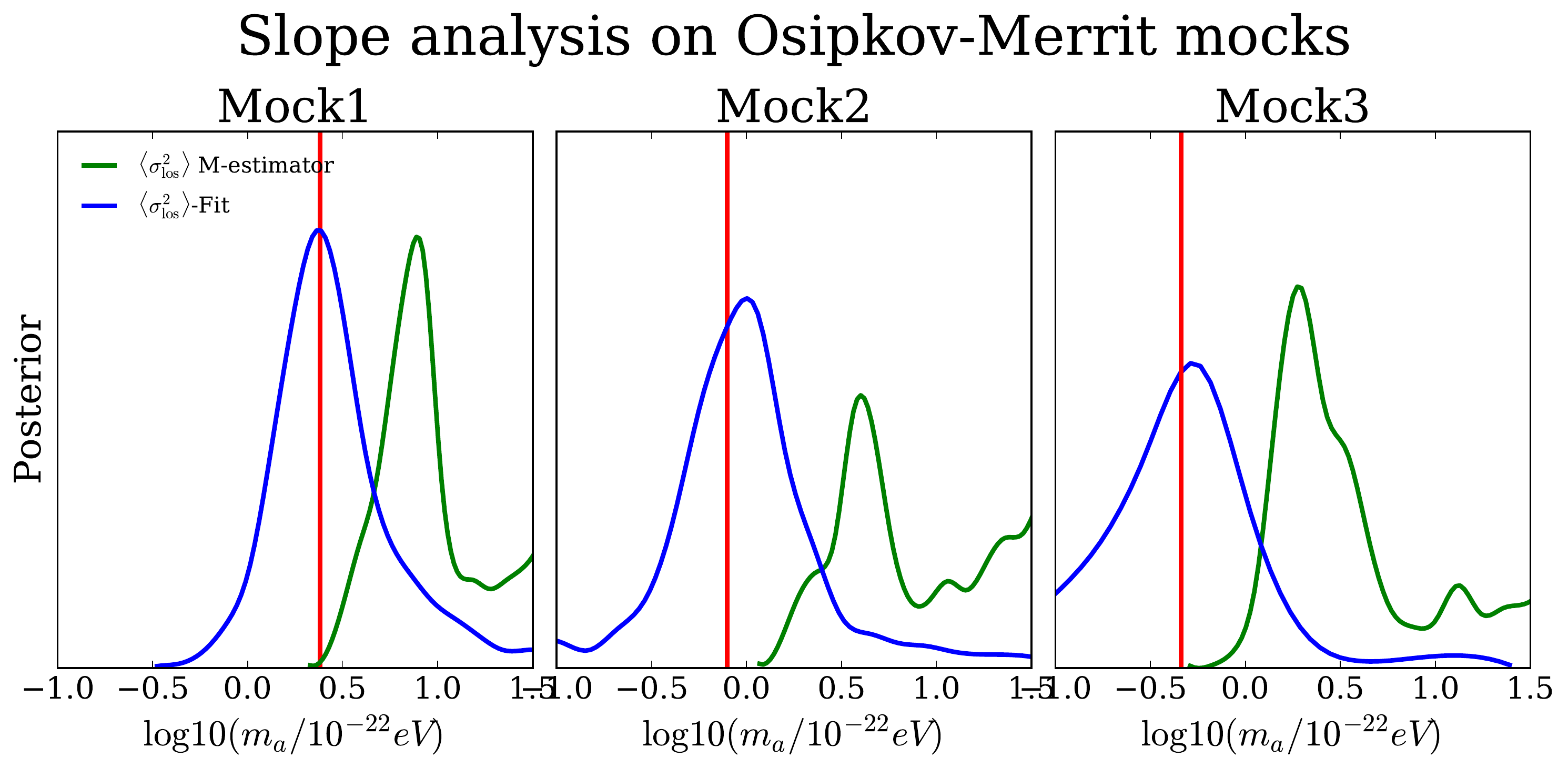}
}

 \caption{\label{mocks}Analysis of three sets of mocks for two stellar
   populations in Fornax and Sculptor-like galaxies, with two
   different considerations for the orbital anisotropy: $\beta=0$
   (left) and the Osipkov-Merrit profile (right). Green contours use the constant virial factor M-estimator, Eq.~\eqref{eq:meanslos_1} with 
$\mu=2/5$, and blue contours use the $\langle\sigma_{\rm los}^2\rangle$-fit, Eq.~\eqref{eq:meanslos_2}. Both estimators recover central axion mass values close to the input (red line), and demonstrate some bimodality when the prior mass range is large (see Figs.~\ref{fig:triangle1} and \ref{fig:triangle} for the full posterior range). The M-estimator has a noticeable bias to larger axion masses, while the
$\langle\sigma_{\rm los}^2\rangle$-fit recovers an unbiased value.}

 \end{center}
\end{figure*}

Results are presented in Fig.~\ref{mocks} where we show the posterior distribution 
for the axion mass for the  three different sets of mocks, in the isotropic (left) and the anisotropic cases (right). 
From left to right, each mock corresponds to a model with larger axion mass, i.e. a smaller core, see
Table~\ref{tab:mocks}. First, notice that the M-estimator (green
lines) tends to over-estimate the axion mass, although the true
value is within the 2-sigma confidence level. We verified that the same conclusion
holds when we do not consider the exterior NFW part of the density
profile. Second, notice that allowing for broad prior ranges (larger than in MP15) the posterior
exhibits a sort of bimodality for the axion mass, which becomes less
relevant as the core size (axion mass) increases (decreases). This
implies that  this method of using two populations will fail to
constrain the axion mass if the core size is much smaller than the
half-light radius of the dwarf. See Appendix \ref{appendix:tplots} for more details. 
Third, we want to stress that our improvement to the method works well independently of the anisotropy of the mocks. 
These results, together with  the information we gather from
Fig.~\ref{fig:slope_WP11}, demonstrate that for the real data we
should fit our model parameters directly to the mean velocity
dispersion for the two populations in order to obtain posteriors with
the least bias.

\subsection{Analysis of Milky Way Dwarf Spheroidals}

\subsubsection{Jeans analysis using the eight classical dSph's}
\label{sec:jeans_data}
\begin{figure}
\includegraphics[width=0.5\textwidth]{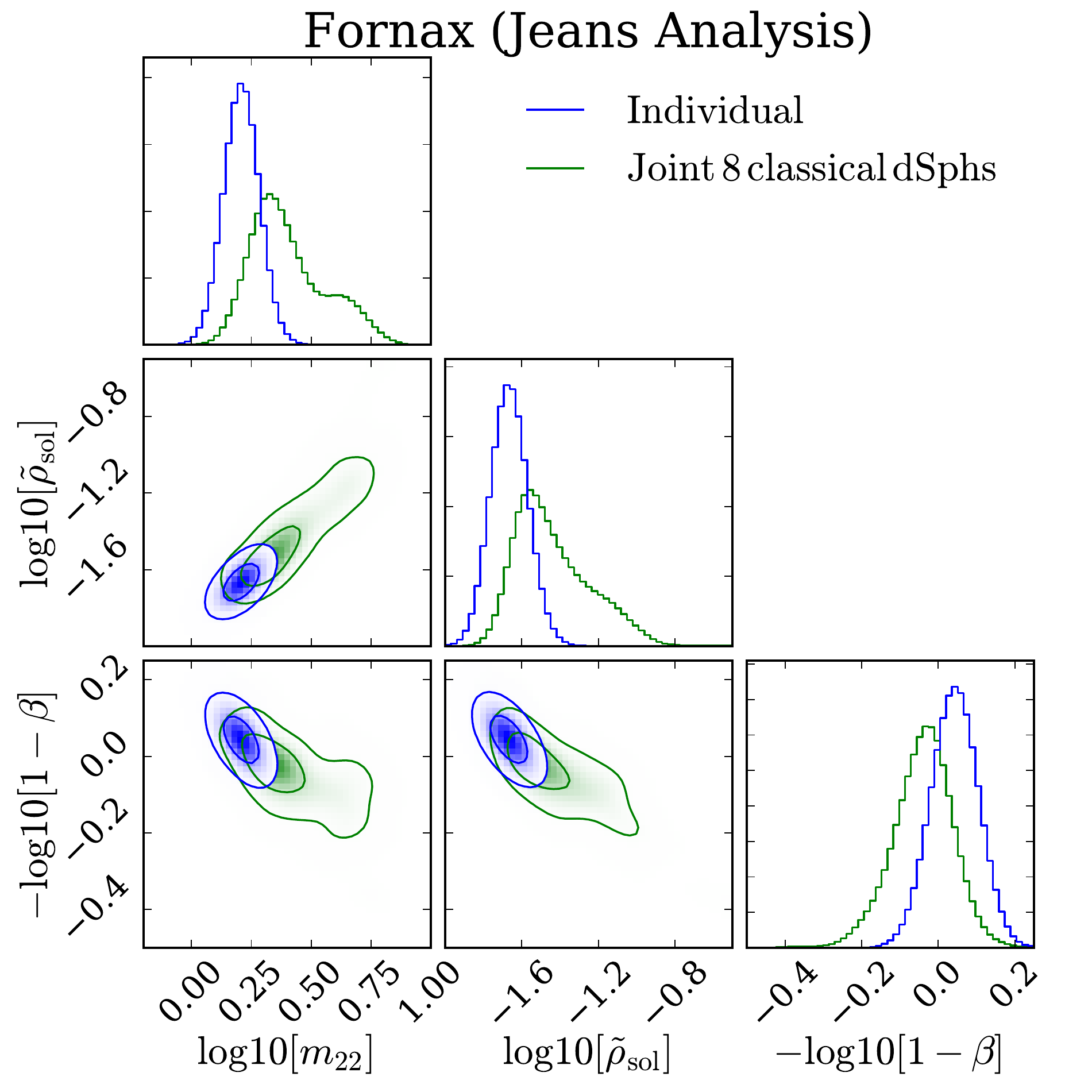}
 \caption{Parameters constrained by the individual Jeans analysis of Fornax, compared to their values in the joint analysis of all eight classical dSphs, where the axion mass is treated as a 
single universal parameter. Contours show 1 and 2-$\sigma$ confidence levels. The joint analysis shifts and broadens the axion mass posterior, caused by a ``compromise'' value between the many 
galaxies. However, due to the $\beta$-degeneracy \emph{the results of
  this analysis should not be read at face value}. axion mass and
soliton density units are $10^{-22} {\rm eV}$ and 
$2.42\times10^{9} M_{\odot} {\rm kpc^{-3}}$, respectively}
\label{fig:fornax_jeans_triangle}
\end{figure} 

Despite the bias introduced by the $\beta$-degeneracy, we present here for completeness the results of Jeans analysis of real stellar dynamical data of the eight classical  dSphs.  We performed two 
such analyses: the first treated each dSph individually (``individual analysis''), and the second treated all eight classical dSphs as a single dataset (``joint analysis''). In the joint analysis, each dSph 
was treated with equal weight in the likelihood function, the axion mass was treated as a universal parameter, and the density profile was taken to have a universal form (effectively assuming that 
stellar feedback plays no significant role in dSphs of different luminosity). For the individual analysis we used the likelihood from Eq.~\eqref{eq:like_jeans}, while for the joint analysis we constructed also a joint likelihood given by the product of the individual likelihoods, one product term for each galaxy. 

For brevity we present the results of these analyses for the halo parameters of Fornax alone. The results are shown in Fig.~\ref{fig:fornax_jeans_triangle}. In the individual analysis, the outer (NFW) 
part of the density profile is unconstrained, and so we show only the parameters constrained by the individual analysis, namely $\{m_a,\rho_{\rm sol},\beta\}$. In the combined analysis the presence 
of smaller and larger dSphs that prefer different values of $m_a$ leads to significant broadening of the posterior distributions compared to the individual analysis.

The individual analysis shown in Fig.~\ref{fig:fornax_jeans_triangle} is consistent with the equivalent analysis (performed while the present work was in preparation) by \citep{2016arXiv160609030C}, 
giving the same central values for $\beta$ and $m_a$. This shows that the methodological differences between the analyses (parameterization of the halos, stellar density profiles, anisotropy model, 
and MCMC methodology) do not lead to significant change of the posterior distributions. 

As in \citep{2016arXiv160609030C}, both our individual and combined analyses show well constrained values of $\beta$, consistent with $\beta=0$. However, this \emph{does not mean that the $
\beta$-degeneracy has been broken}. The limit of the axion density
profile to NFW as $m_a\rightarrow\infty$ does not allow the Jeans analysis to distinguish cusped from cored profiles, as we 
demonstrated using mock data in Figs.~\ref{fig:scl_} and \ref{fig:for_}. Due to the $\beta$-degeneracy, Jeans analysis returns  small mass values, and large core sizes, even in the case that the NFW 
profile is the correct one. The results in Fig.~\ref{fig:fornax_jeans_triangle} should not be read at face value as constraints on the axion mass or core size.

\begin{figure}
\begin{center}
{\includegraphics[width=0.495\textwidth]{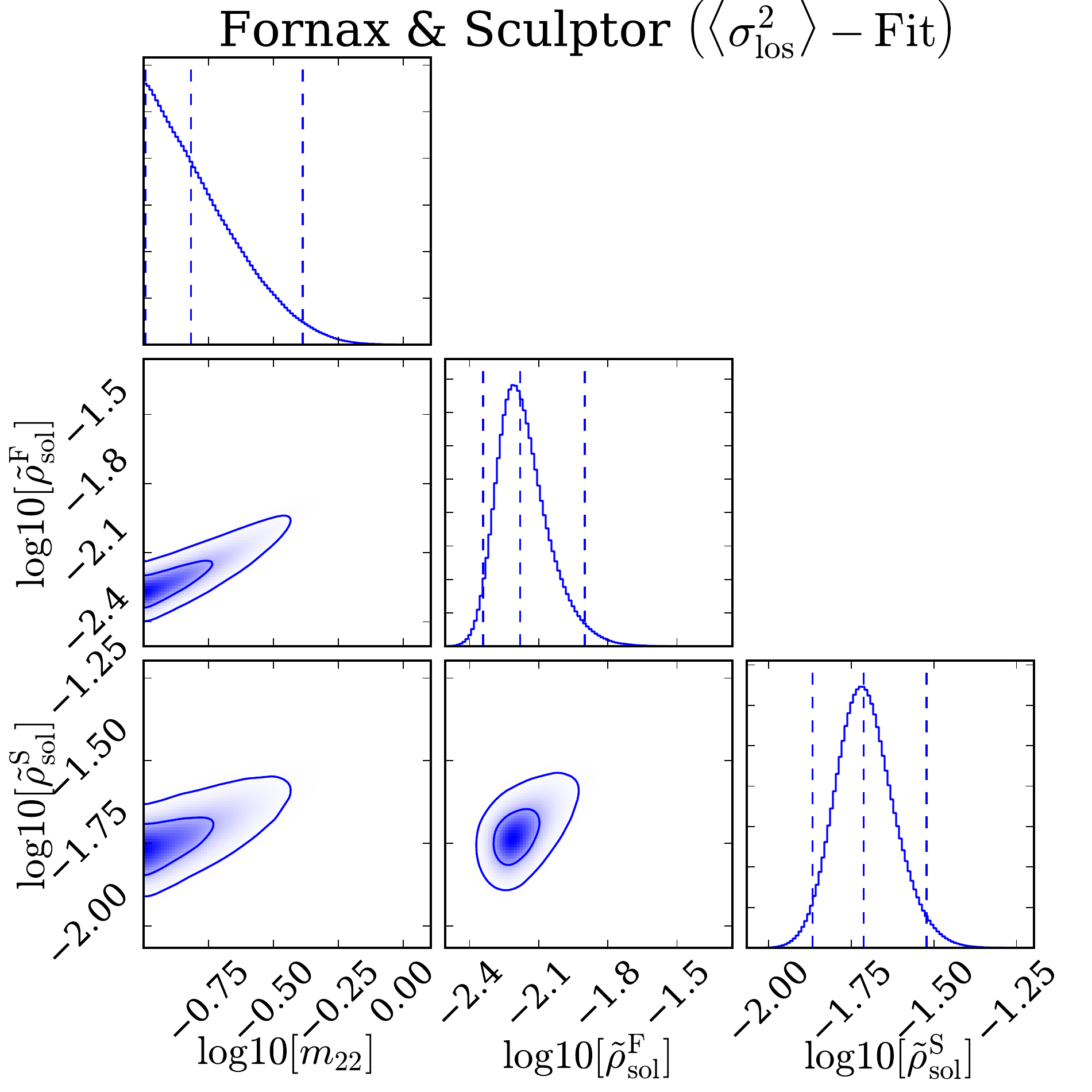} }
 \caption{Posterior distribution of the central density and axion mass
   for the joint analysis of Fornax and Sculptor using the
   $\langle\sigma_{\rm los}^2\rangle$-fit. Compared to MP15 (which
   used the M-estimator, see Fig.~\ref{fig:mass-lik}), using the
   $\langle \sigma_{\rm los}^2\rangle$-fit leads us to find a
   significant shift in the inferred value of axion mass towards
   smaller values with a \emph{very tight} bound: $m_a<0.4\times
   10^{-22}\text{ eV}$ (97.5\% C.L.).The reason for the shift in the
   inferred  axion mass  between the two estimators is explained in
   Fig.~\ref{fig:slope_WP11} by the non-constancy of the virial factor
   with respect to the model parameters. Contours show 1 and
   2-$\sigma$ confidence levels. axion mass is given in units of
   $10^{-22} {\rm eV}$ and soliton density in units of $2.42\times10^{9} M_{\odot} {\rm kpc^{-3}}$.}
 \label{fig:slope_data}
 \end{center}
\end{figure}

\subsubsection{Averaged velocity dispersion of Fornax and Sculptor}
Here we use again the likelihood defined in Eq.~\eqref{eq:like_slope}, with the data obtained from WP11 for Fornax and Sculptor galaxies. Fig.~\ref{fig:slope_data} shows the results for our  fit using the unbiased $\langle\sigma_{\rm los}^2\rangle$-fit, in the joint analysis of Fornax and Scultor. We also performed this analysis using the M-estimator (not shown) and found results consistent with those of MP15. The internal kinematics of Fornax and Sculptor  only give an upper limit on the axion mass, giving a constraint of 
$m_a<0.4\times 10^{-22}\text{ eV}$ (97.5\% C.L.) using the $\langle\sigma^2_{\rm los}\rangle$-fit. As expected based on our analysis of the mocks, the new constraint is shifted towards smaller 
values of the axion mass and central densities compared to the M-estimator used in MP15. The shift in central density is understood as a consequence of the non-constancy of the virial factor with 
respect to the model parameters, as demonstrated in Fig.~\ref{fig:slope_WP11}. 

As also anticipated form the mocks, the limits we find are slightly sensitive to the lower bound of the axion mass prior. However, even considering a 3-sigma level for the upper limit, the $\langle
\sigma^2_{\rm los}\rangle$-fit still finds significantly tighter bounds on $m_a$ than MP15. Note also that for this analysis we kept only the soliton part of the model, for which the relevant parameters 
are $m_{22}$ and $\rho_{\rm sol}$, since the exploration with mocks showed that the rest of the free parameters in Eq.~\eqref{eq:density1} are totally unconstrained. 

It is important to stress again that our mocks indicated that the $\langle\sigma_{\rm los}^2\rangle$-fit is unbiased with respect to the axion mass using the method of WP11. The mocks further 
demonstrate that if the true axion mass were $m_a=2.4\times 10^{-22}\text{ eV}$, as inferred by the combined Jeans analysis, then the $\langle\sigma_{\rm los}^2\rangle$-fit should recover this well. 
Since this was not the case, and we recovered an upper limit inconsistent with the Jeans analysis (see Fig.~\ref{fig:mass-lik}), we must infer that the constraint on the axion mass from Jeans analysis 
of dSphs is significantly biased to incorrect values due to the $\beta$-degeneracy.

\section{Discussion}
\label{sec:cosmology}
The constraint we find in section \ref{sec:results} requires axion DM to be extremely light, which has very important consequences for cosmology. Axion DM suppresses structure formation relative to 
CDM, and if it is too light it may be in conflict with observations~\citep[e.g.][]{khlopov_scalar,2010PhRvD..82j3528M}. 

\subsection{The Number of MW Satellites}

In order to assess whether our constraint is consistent with cosmological models, we perform two simple estimates of the subhalo mass function based on the axion linear mass power 
spectrum \citep[we use \textsc{axionCAMB}][]{Hlozek:2014lca,2016arXiv160708208H,camb};\footnote{\textsc{axionCAMB} produces equivalent results to the \textsc{class} \citep{2011JCAP...07..034B} based code of \citep{2016JCAP...07..048U}, and at low redshifts also matches the analytic transfer function of \citep{2000PhRvL..85.1158H}.} and the halo mass function of \citep{1999MNRAS.308..119S}. The first estimate uses the scale-dependent barrier for axions of \cite{2014MNRAS.437.2652M}, the approximate progenitor mass function of \citep{2004ApJ...609..474B}, and the subhalo mass function of \citep{2008MNRAS.387..689G}. The second is based on the method in \cite{2016JCAP...04..059S,2017arXiv170407838M}.\footnote{We thank Riccardo Murgia for providing the code for the second calculation.}

These calculations return an un-normalized mass function for subhalos of mass $M$ in a parent halo of mass $M_0$, $n_{\rm sub}(>M|M_0)$. For $M_0=10^{12}M_\odot\approx M_{\rm MW}$, where $M_{\rm MW}$ is the mass of the Milky Way, the ULA subhalo mass function is equivalent to CDM for large $M$. Therefore, we normalize our mass function by applying a constant multiplicative factor such that the CDM mass function computed in this manner matches $N$-body simulation results. We normalize to 
\cite{2014MNRAS.439..300L}, who found $n_{\rm sub}(>10^{10}M_\odot|M_{\rm MW})\approx 10$ for CDM, and apply the same normalization factor to the ULA models. 

In both calculations we find that ULAs consistent with corrected slopes analysis produce far too little substructure to be consistent with the observations: $m_{22}
<0.4$ cannot even give the eight classical dSphs, never mind passing a more realistic bound such as $n_{\rm sub.}\gtrsim 66$~\citep{2014MNRAS.439..300L}. The subhalo mass function for $m_{22}<0.4$ turns over and produces no subhalos with $M\gtrsim 10^{9-10}$, which is in conflict with substructure bounds from tidal streams~\citep{2017MNRAS.466..628B}. These observations indicate that ULAs may suffer 
from a similar \emph{Catch 22} to Warm Dark Matter (WDM): if you want cores, you don't get enough satellites; if you want enough satellites, you don't get big enough cores. The two calculations of the subhalo mass function we have considered differ by a factor of a few for larger ULA masses, pointing to the need for improved calculations and simulations to derive accurate conclusions.

The full excursion set calculation for the subhalo mass function with axion DM was recently performed by \cite{2017MNRAS.465..941D}. The excursion set calculation of \cite{2017MNRAS.465..941D} leads 
to a \emph{larger} cut-off scale on the halo mass function for fixed axion mass, with the qualitative effects being consistent with our approximate treatments. Thus in the full calculation there will be even less 
substructure than we have estimated, strengthening our conclusion that ULAs able to provide large cores to Fornax and Sculptor will struggle to produce the observed number of MW satellites (they 
``over-solve'' the ``missing satellites'' problem). The inability of the subhalo mass function to produce enough satellites does not necessarily exclude the ULA model, however: the mass function gives the 
expected number, and it could be that the MW is an outlier. There are analytic tools available that use extreme value statistics to compute exclusions based on the most massive objects~\citep[e.g.]
[and references therein]{2011MNRAS.413.2087D}, and these could be extended to the lower end of the mass function.

Ultimately, the ULA subhalo mass function should be obtained in MW zoom-in simulations, as was done for WDM by e.g. \citep{2014MNRAS.439..300L,2016MNRAS.459.1489B}. Codes capable of 
such simulations of ULAs have now been developed by \cite{2016PhRvD..94d3513S} and \cite{2016PhRvD..94l3523V}. Simulations will quantify the scatter around (semi-) analytic results for the mass function, such as Press-Schechter. Zoom-in simulations will also allow us to study the effect of tidal disruption, which  
for cored density profiles might be as important as an initial cut-off on the mass power spectrum. This has been recently investigated in \citep{2017MNRAS.465L..59E}, where they analyze how the 
number of substructures, in a Milky Way like galaxy, is affected when falling into the disc potential, depending on their initial density profile. They find that the survival of cuspy satellites is almost 
twice larger than the cored ones. For the axion model, the inverse relation between the central density and the core radius, could  lead to a different conclusion, but, once again, simulations are 
required to say how axion DM halos will respond to the dynamical interaction with baryons. The ``quantum'' stripping discussed in \cite{2017PhRvD..95d3541H} will also play an important role.

\subsection{Other Constraints}
Here we discuss wider implications of our finding, $m_{22}<0.4$, showing how the dSph stellar kinematics are complementary to other probes of DM.  The first observable we want to discuss is CMB, 
which has been thoroughly studied in ULAs models. Taken at face value, however see section \ref{sec:feedback}, our bound is consistent with the current constraints from precision 
cosmology, in the form of the \emph{Planck} temperature power spectrum, which requires $m_a\gtrsim 10^{-24}\text{ eV}$~\citep{Hlozek:2014lca}, and it may well be tested to a higher precision in 
the near future by the lensing power spectrum measured by CMB polarisation Stage-IV ground based telescopes~\citep{2016arXiv160708208H}.

Interestingly, our inferred mass limit is consistent with the
interpretation of ULA quantum pressure as the origin of the offset
between dark and ordinary matter in Abell 3827, which requires $m_a
\approx 2\times 10^{-24}\text{eV}$~\citep{2016PDU....12...50P}. Our
bound is also consistent with constraints based on the survival of the
cold clump in Ursa Minor and distribution of globular clusters in
Fornax, which require $m_a\sim 0.3 - 1\times 10^{-22}\text{
  eV}$~\citep{2012JCAP...02..011L}. On the other hand, our bound is in
$\gtrsim 2 \sigma$ tension with the earlier Jeans analysis of Fornax
by \citep{2014NatPh..10..496S}, $m_a=8.1^{+1.6}_{-1.7}\times
10^{-23}\text{ eV}$ (though for the reasons explained, Jeans analysis
in this case leads to biased results). Explaining the half-light mass
in the ultra-faint dwarfs requires a somewhat larger ULA mass of
$m_a\sim 3.7 - 5.6\times 10^{-22}\text{ eV}$
\citep{2016MNRAS.460.4397C}, which is also in tension with our bound,
as is the estimation $m_a \sim 10^{-21}\text{ eV}$
in\citep{Urena-Lopez:2017tob}, though the errors are large and hard to
fully quantify. 

Next, we want to compare with constraints from reionization. In this case the comparison is more complicated because the physics of the baryons plays a very important role. Establishing such 
bounds rigorously, in any given model of DM,  requires dedicated studies of the evolution of the mass power spectrum in the quasi-linear regime and modeling of the intergalactic medium (IGM). 
Some work along these lines has been done by \citep{2015MNRAS.450..209B} and \citep{2016JCAP...04..012S}. The limit on ULA mass from our analysis of the dSph data, $m_{22}<0.4$, gives a 
considerable small value for the CMB optical depth, $\tau$, which is  in tension with the Planck+ Low-l WMAP 9 (Planck + WP)  constraints on $\tau$, yet this is consistent with the Planck High 
Frequency Instrument  (Planck+HFI) constraint. This demonstrates the power that future constraints on the epoch of reionization from CMB polarization will have to probe the nature of DM \citep[e.g.]
[]{2014JCAP...08..010C}, and the importance of understanding possible low-$\ell$ polarization systematic errors that could be causing a tension between \emph{Planck} and WMAP $\tau$ 
measurements. See appendix \ref{sec:tau} for wider explanation on how we derived these constraints  based on \cite{2015MNRAS.450..209B}. 

 \citep{2016JCAP...04..012S} finds that  $m_a>2.6\times 10^{-23}\text{ eV}$ is consistent with their  reionization model based on $N$-body simulations demanding an ionized 
fraction of HI of 50\% by $z=8$ ($m_a>10^{-23}\text{ eV}$ from collapsed mass fraction inferred from Lyman-$\alpha$ absorbers).  Our bound is also inconsistent with the Hubble Ultra Deep Field UV-Luminosity function \citep{2015ApJ...803...34B}, and a compilation of UV-luminosiy function data~\citep{2017PhRvD..95h3512C}, which both require $m_a\gtrsim 10^{-22}\text{ eV}$.

The strongest constraints on any possible suppression of clustering power relative to CDM are found from the Lyman-$\alpha$ forest flux power spectrum. Recently, two groups have performed Lyman-$\alpha$ simulations with ULAs, finding the constraints $m_{22}\gtrsim 20-30$, depending on the exact combination of data used~\citep{2017arXiv170309126A,2017arXiv170304683I}. It is important to note, however, that Lyman-$\alpha$ constraints depend strongly on the modelling of the temperature of the intergalactic medium, and in particular can be loosened if the evolution is non-monotonic~\citep{2015arXiv151007006G,2017PhRvD..95d3541H}.

\subsection{The role of feedback and measurements on other dSphs}
\label{sec:feedback}
The main assumptions in this paper is that there is a universal density profile, and that the axion mass can be treated as universal parameter. Such conditions  are  only consistent if processes  
associated to the presence of a barionic component, stellar feedback particularly, affects dSph density profiles minimally. Unfortunately, as far as we know, there do not exist simulations of the axion 
model studying how feedback acts on scalar field halos like those we study. 

For the sake of discussion, imagine separating the axion halo into the soliton, which responds as a coherent field, and the NFW piece, which is incoherent and responds as CDM. What we can guess 
about the effects of feedback based on existing work depends on how big the core of the galaxy is.  If the actual axion mass is large, the ``true’’ core is small, and feedback will act on the outer part 
of the halo, creating a ``false’’ core out of this initially NFW-like piece, if this external part actually behaves as CDM. In this case, just as for CDM, dSph density profiles will not be universal, and the 
(luminosity-dependent) effects of feedback will be seen\citep{2016MNRAS.459.2573R,2016MNRAS.457.1931S,2014Natur.506..171P,2012MNRAS.422.1231G,2010MNRAS.402.1599S}. 
However, if the axion mass is small, and is itself responsible for the large cores in dSphs, then feedback will act mostly on the soliton piece. In this case, because the soliton is the ground-state of 
axion DM, the halo relaxes back to the universal profile under perturbations. Numerical simulations indicate that the relaxation time, even for strong perturbations, is of order $t_{\rm rel.}\sim 10^3/
m_a\sim 10^3 m_{22}^{-1} \text{ years}$~\citep{2006ApJ...645..814G}.\footnote{Rapid relaxation is also observed in violent events, such as multi-soliton mergers~\citep{2016PhRvD..94d3513S}.} 
According to these studies, the relaxation time for our benchmark axion mass is vanishingly small in astrophysical terms. As long as the last feedback event (e.g. supernova explosion) of relevance 
occurred some time $t_{\rm feed.}>t_{\rm rel.}$ ago, then the soliton will have relaxed back to the groundstate, and the measured $(\rho_{\rm sol.},r_{\rm sol.})$  can be used to reliably infer the axion 
mass. 

The above argument suggests a powerful probe to test the axion model as an explanation of dSph cores versus the need for feedback. If the axion explanation for cores is correct, then our discussion 
of relaxation times suggests that feedback cannot affect the universal nature of the density profile. Using multiple dSphs, each with multiple stellar populations, this universal nature can be tested. The 
current constraints from Fornax and Sculptor cannot be used to infer whether the density profile is universal: both galaxies are cored and the size of the core is not bounded from above, and we thus 
only obtain an upper limit on the axion mass. However, by measuring multiple populations in more dSphs, it may be possible in future to test the universal profile. If future measurements do not find a 
universal profile, or find inconsistent limits on the axion mass (for example, a large lower limit from a faint, cuspy, dSph), then axions cannot be the sole explanation for dSph cores, and feedback (or 
some other new DM physics) must be operative. From our tests with mocks (not shown here) we think that the ideal scenario would be a galaxy where two populations are identified and they have such 
half mass radius that one is very well embedded in the soliton (core), and the other probes the outer part of the DM halo. In this way we would be able to find an axion mass bounded from above and 
below.

\section{Conclusions}
\label{sec:conclusions}

We used mocks of dSphs embedded in an axion DM halo to test for the presence of bias in constraints to the particle mass in axion DM models. The main points  to conclude are: 
\begin{itemize}
\item Using Jeans analysis with constant unknown anisotropy to fit the line of sight velocity dispersion leads to biased constraints on the axion mass, to the point where one can conclude that  
galaxies are well fitted by the axion halo model (cored) when in reality the underlying model is a "cuspy" one.
\item We also found that using the M-estimator to fit the slope defined by the  mean velocity dispersion from two different  stellar populations in the same galaxy also leads to biased constraints, 
though to a lesser extent. As expected, the bias is worse when the axion mass is smaller since this case corresponds to very large cores.   
\item An intermediate approach  where we compute the mean velocity dispersion from direct integration of the Jeans equation, Eq.~\eqref{eq:meanslos_2},  and fit the luminosity averaged velocity 
dispersion of two stellar subpopulations seems to  provide unbiased constraints on the halo parameters.
\end{itemize} 

Fitting $\langle\sigma_{\rm los}^2\rangle$ rather than using the M-estimator (used by MP15) in the joint analysis for data from Fornax and Sculptor galaxies taken from WP11 leads to a tighter limit to the axion mass,  
$m_a<0.4\times 10^{-22}\text{ eV}$. We have corroborated that our new approach works very well independently of the true anisotropy profile, by applying it to isotropic and anisotropic mock data.  This is clearly in tension with a ``blind'' Jeans analysis, performed using a joint likelihood of the eight classical dSph's, with free constant anisotropy parameter $
\beta$, which gives $m_a=2.4^{+1.3}_{-0.6}\times 10^{-22}\text{
  eV}$. However, our analysis of mocks leads us to believe that the
estimate of $m_a$ from Jeans analysis suffers from significant bias, see Fig.~\ref{fig:mass-lik} for a comparison. Without proper knowledge of the dSph velocity anisotropy it is not possible to extract unbiased constraints on DM models. Here are two important points to 
remark. First, we are not attempting to extract more information from the other dSphs because we cannot be certain that those galaxies actually have a cored density profile. Second, we have verified 
that constraints set individually from Fornax and Sculptor are compatible, otherwise stating a joint likelihood would not be correct. From the $\langle\sigma_{\rm los}^2\rangle$ individual fit we 
obtained $m_{22}<0.48$ and$m_{22}<0.79$ for Fornax and Sculptor respectively. 

The tight bound on $m_a<0.4\times 10^{-22}\text{ eV}$ required if ULAs are to provide kpc-scale cores to Fornax and Sculptor (WP11 cores) runs into several problems when faced with cosmology. 
Firstly, we performed a simple estimate of the subhalo mass function, and demonstrated that such a ULA cannot provide enough dwarf satellites of the MW to be consistent with 
observations. This suggest that ULAs, like WDM, may suffer from a \emph{Catch 22} in that ``if you want large cores, you don't get enough dwarfs; if you want enough dwarfs, you don't get big 
enough cores''. Existing cosmological bounds also appear to rule out a ULA origin for the WP11 cores: high-$z$ galaxies rule out $m_a\lesssim 1\times 10^{-22}\text{ eV}$~\citep{2015MNRAS.450..209B,2016ApJ...818...89S}. 
Constraints from reionization, such as the CMB optical depth $\tau$, are less conclusive, and may even favour low axion mass. If ULAs satisfying our bound are 
indeed the DM, they can be definitively ruled out either by proper analysis of existing Lyman-$\alpha$ forest power spectrum data~\citep{2017arXiv170309126A}, or by upcoming CMB Stage-IV lensing power 
spectrum measurements~\citep{2016arXiv160708208H}. However, we would like to emphasis the need of detailed studies of the interplay between the physics of the axion DM and the physics of baryons, in  order to make more consistent comparisons.

Improvements on DM constraints from stellar kinematics may be obtained from future measurements of proper motions \citep[e.g.][]{2007ApJ...657L...1S}. Extending our present analysis 
(chemodynamical modelling of multiple stellar populations combined with the $\langle\sigma_{\rm los}^2\rangle$-fit) to other dwarfs beyond Fornax and Sculptor can be achieved by multi-object 
spectrographs attached to several-metre telescopes. On a short time scale we have WEAVE\citep{2014SPIE.9147E..0LD}, MOONS \citep{2012SPIE.8446E..0SC} and 4MOST \citep{2012SPIE.8446E..0TD}. Using these data and our methodology could significantly improve our ability to test models of DM using stellar dynamics. 

\section*{Acknowledgments}

We are grateful to Matthew Walker for providing the data, to Renan Barkana, Erminia Calabrese, Malcolm Fairbairn, Carlo Giocoli for useful discussions, to Ren\'{e}e Hlozek for providing a \textsc{python} 
implementation of the spectral convergence test of \citep{2005MNRAS.356..925D}, to Brandon Bozek for collating the results of \citep{2015MNRAS.450..209B}, and to Riccardo Murgia for providing code for the second subhalo mass function calculation.  DJEM acknowledges the support 
of a Royal Astronomical Society research fellowship, hosted at King's College London.  AXGM acknowledges support from C\'atedras CONACYT and UCMEXUS-CONACYT collaborative project 
funding. This work was partially supported by PRODEP, DAIP-UGTO
research grant 732/2016 and 878/2016, PIFI, CONACyT M\'exico under
grants 232893 (sabbatical), 182445, 167335, 179881, 269652 and 
Fronteras 281, Fundaci\'on Marcos Moshinsky, and the Instituto Avanzado de Cosmolog\'ia (IAC) collaboration. 

\bibliography{ula_dsph}
\bibliographystyle{mnras}

\appendix
\section{Jeans Analysis}
\label{appendix:jeans}

Following the standard parametric analysis analysis~\citep{2009ApJ...704.1274W, Salucci:2011ee}  we will consider that the stellar component in each individual galaxy is in dynamical equilibrium 
and that stars are kinematic tracers of the underlying DM potential . Assuming, further, spherical symmetry, Jeans's equation relates the mass profile of the DM halo, 
\begin{equation}
M(r)=4\pi \int_0^r{\rho(r') r^{'2} }dr'
\end{equation}
to the first moment of the stellar distribution function, 
\begin{equation}
\frac{1}{\nu}\frac{d}{dr}\left(\nu\langle v^2_r\rangle\right)+2 \frac{\beta\langle v^2_r\rangle}{r} = -\frac{G M}{r^2}\,.
\end{equation}
Above, $\nu(r)$, $\langle v^2_r (r)\rangle$, and $\beta(r)=1-\langle v^2_{\theta}\rangle/\langle v^2_r\rangle$ are the three-dimensional 
density, radial velocity dispersion, and orbital anisotropy, respectively, of the stellar component.
The parameter $\beta$ quantifies the degree of radial stellar anisotropy:  
if all orbits are circular $\langle v^2_r\rangle = 0$, and then $\beta = \infty$; if the orbits are isotropic 
$\langle v^2_r\rangle=\langle v^2_{\theta}\rangle$, and $\beta= 0$; finally, if all orbits are perfectly radial, 
$\langle v^2_{\theta}\rangle = 0$, then $\beta = 1$. There is no preference {\it a priori} for either radially, $\beta >0$, or tangentially, $\beta<0$, biased systems.

The function $F$ depending on the anisotropy is given by:

\begin{equation}
F(\beta,R,r')\equiv \int_R^{r'}{dr \left(1-\beta\frac{R^2}{r^2}\right) \frac{r^{-2\beta +1}}{\sqrt{r^2-R^2}}}\,,
\end{equation}
For the stellar density we adopted a  Plummer profile,
\begin{equation}\label{eq.Plummer}
I(R)=\frac{L}{\pi R_{\rm half}^2}\frac{1}{[1+(R/R_{\rm half})^2]^2}\,,
\end{equation}
where $L$ is the total luminosity of the object and $R_{\rm half}$, the half-light radius. The values of these two quantities for each of the eight classical dSphs are listed in Table~I of \citep{2009ApJ...704.1274W}.
Under the assumption of spherical symmetry the corresponding three-dimensional stellar density associated with the  Plummer profile takes the form 
\begin{equation}
\nu(r)=\frac{3 L}{4\pi r_{\rm half}^3}\frac{1}{[1+(r/r_{\rm half})^2]^{5/2}}\,.
\end{equation}

For the Jeans analysis we adopted the following priors: 

\begin{subequations}
\begin{eqnarray}
-3.0<&\log10 \left( \frac{m_{a}}{10^{-22} {\rm eV}}\right)&<5.0 \,, \\
-3.0<&\log10 \left(\frac{\rho_{\rm sol,i}}{2.42\times 10^9 M_{\odot}\rm kpc^{-3}}\right)&<3.0 \,, \\
-5<&\ln \left(\epsilon_{i}\right)&<0 \,, \\
-3<&\ln \left(\frac{r_{s,i}}{\rm kpc }\right)&<3 \,, \\
-3<&-\ln \left(1-\beta_{i}\right)&<5. \,.
\end{eqnarray}
\end{subequations}

\section{Triangle plots of mock galaxies}
\label{appendix:tplots}
\begin{figure*}
\begin{center}
\includegraphics[width=0.32\textwidth]{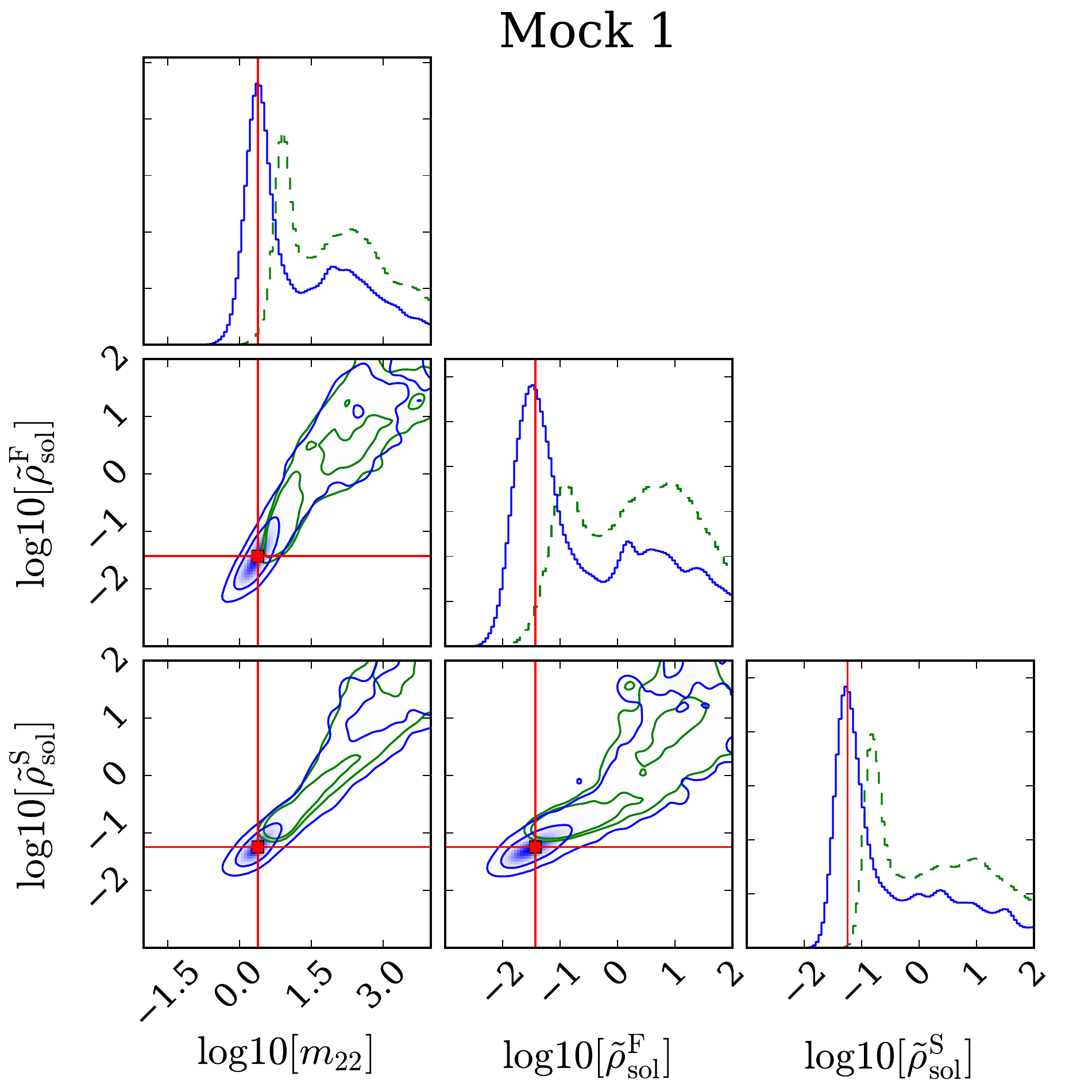} 
\includegraphics[width=0.32\textwidth]{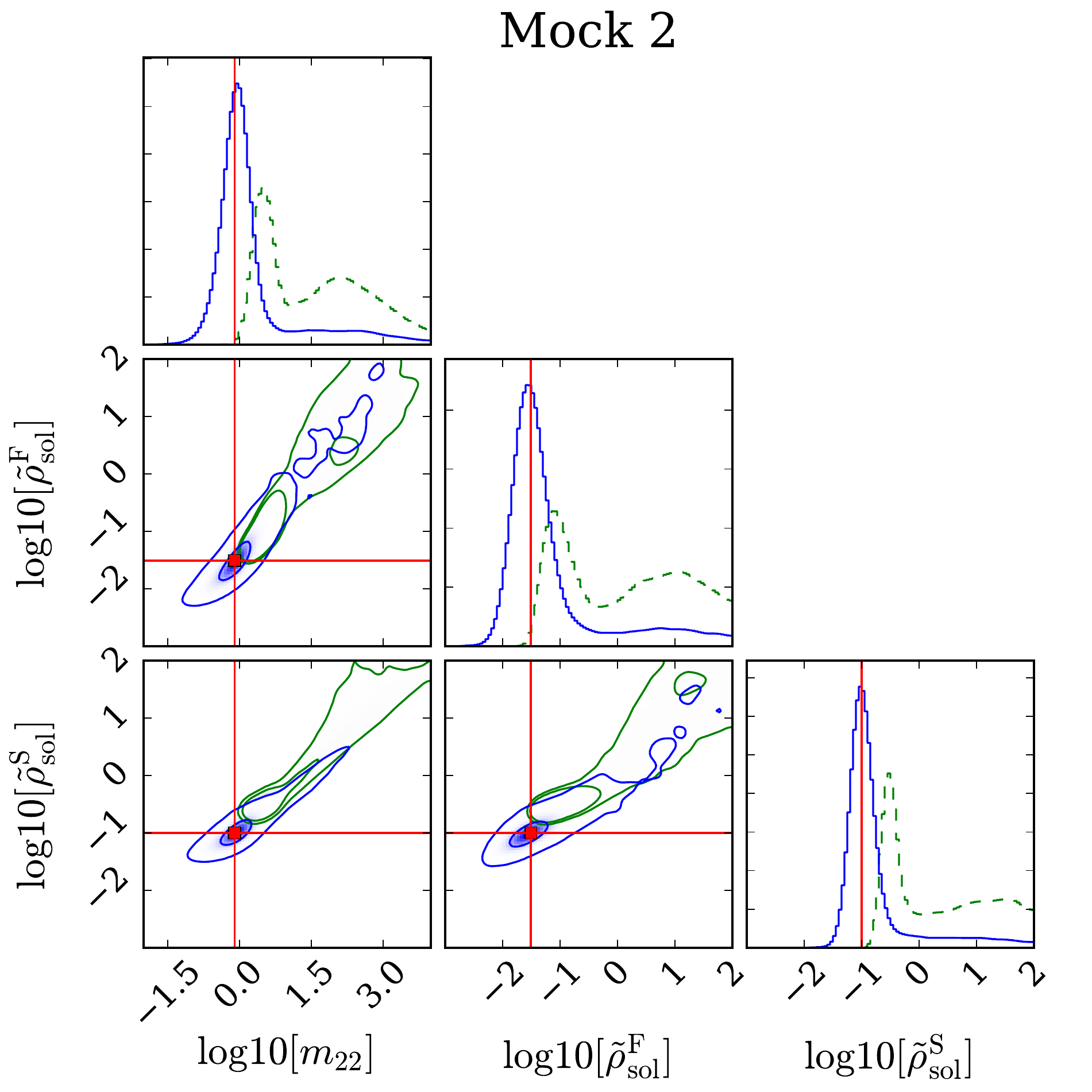}
\includegraphics[width=0.32\textwidth]{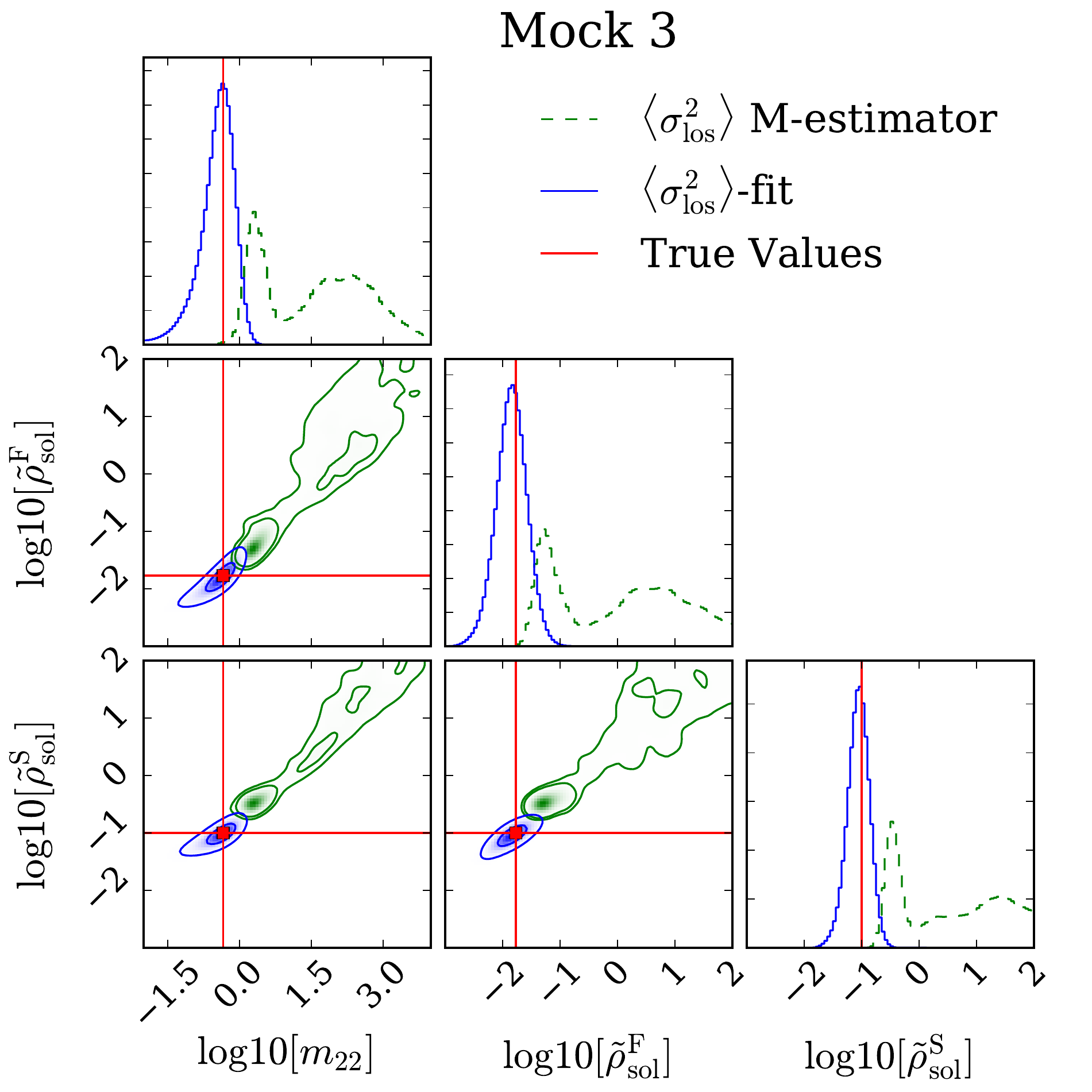} 

\caption{\label{fig:triangle1}Full analysis of three sets of mocks
  already presented in Fig~\ref{mocks}, but only for the particular
  case of $\beta=0$. Green contours, dashed lines, use the constant virial factor M-estimator, Eq.~\eqref{eq:meanslos_1} with 
$\mu=2/5$,  and blue contours use the $\langle\sigma_{\rm
  los}^2\rangle$-fit, Eq.~\eqref{eq:meanslos_2}. As explained before,
both estimators recover central axion mass values close to the input
(red line), and some bimodality can be clearly seen when the prior
mass range is large. The M-estimator has a bias to larger axion
masses, while the $\langle\sigma_{\rm los}^2\rangle$-fit recovers an
unbiased value. Contours indicate the 1 and 2-$\sigma$ confidence
levels. axion mass and soliton density are given in units $10^{-22}
{\rm eV}$ and $2.42\times10^{9} M_{\odot} {\rm kpc^{-3}}$
respectively.}
 \end{center}
\end{figure*}

\begin{figure*}
\begin{center}
{\includegraphics[width=0.32\textwidth]{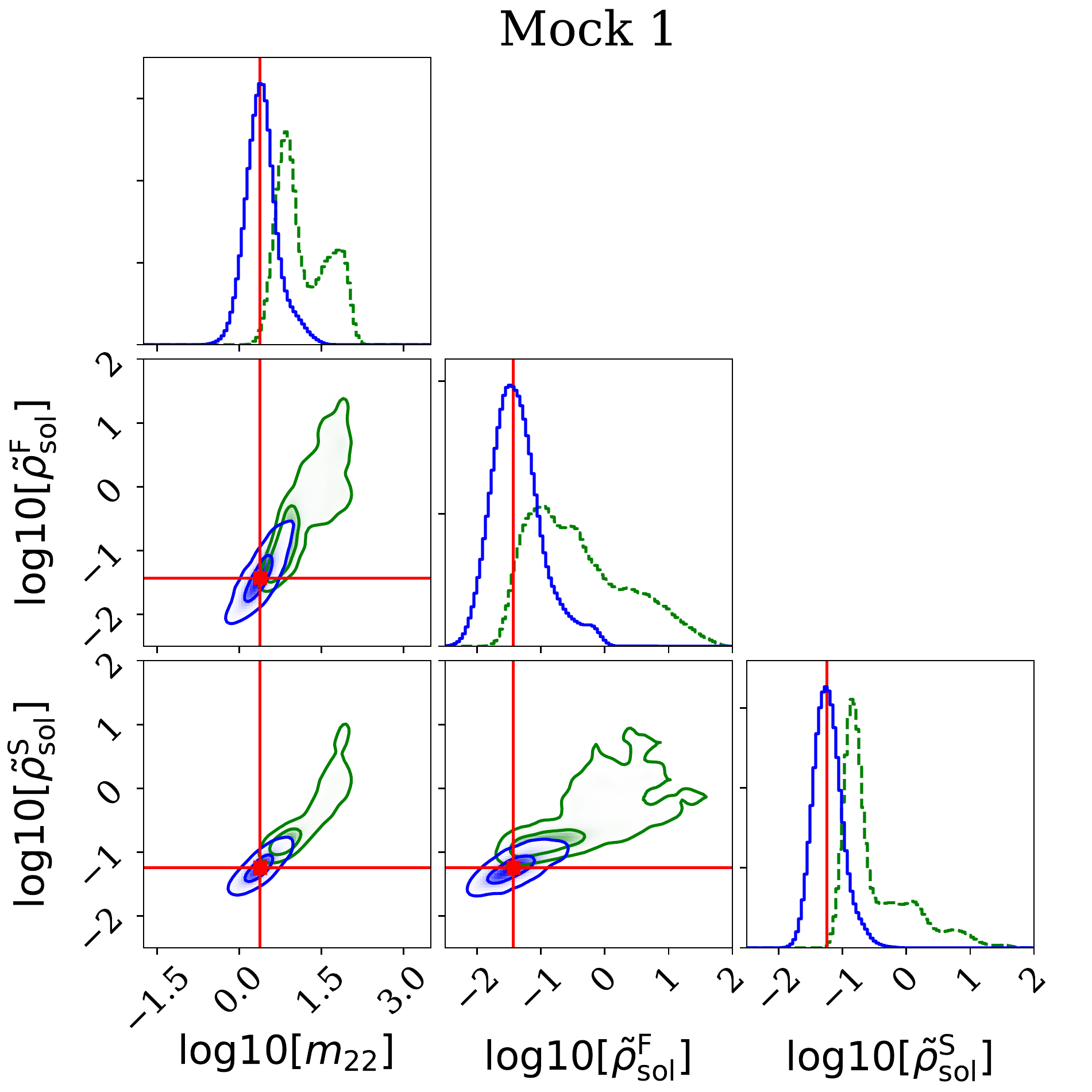} 
\includegraphics[width=0.32\textwidth]{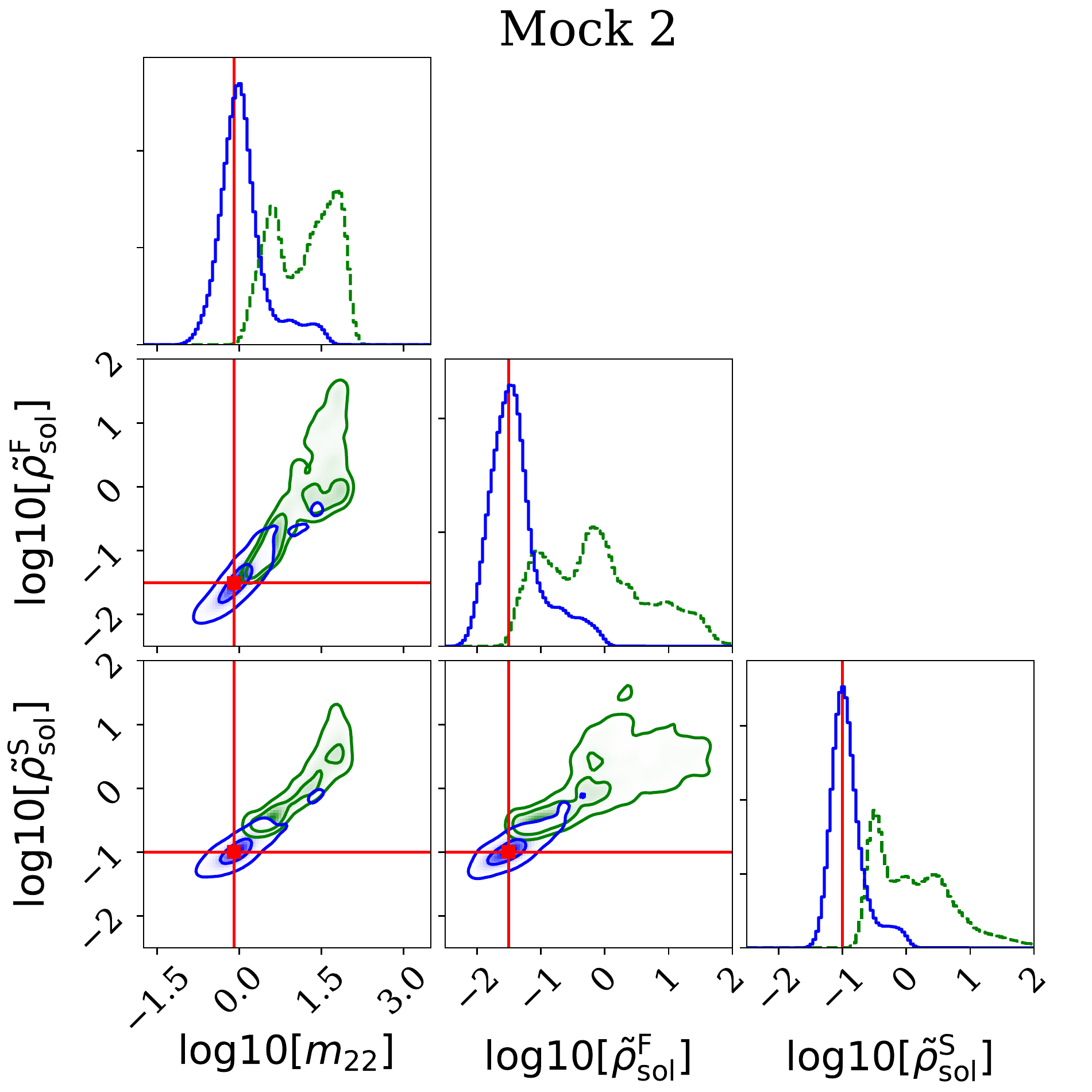}
\includegraphics[width=0.32\textwidth]{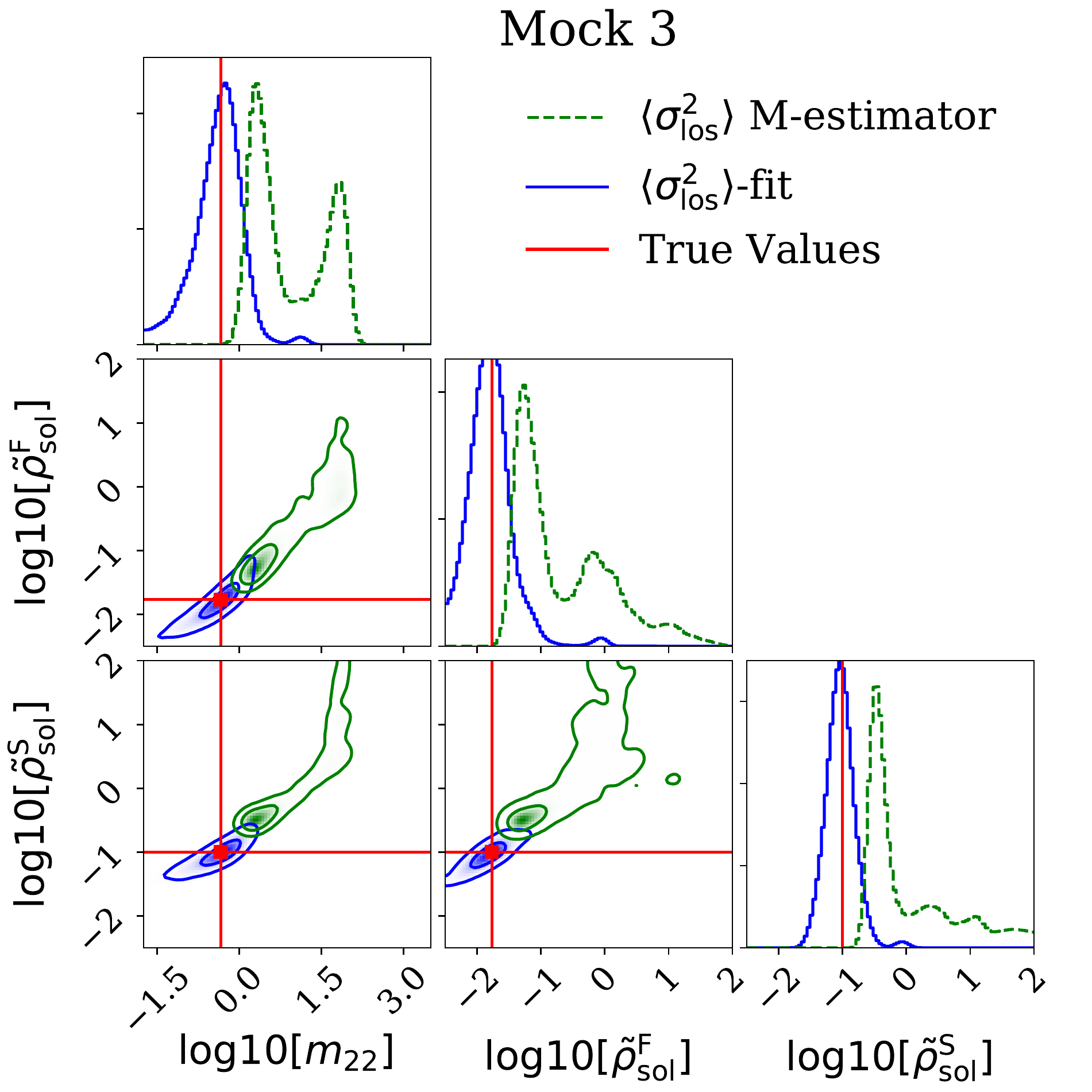} 
}
\caption{\label{fig:triangle} Same as Fig.~ \ref{fig:triangle1} but for the set of mocks constructed with an Osipkov-Merrit anisotropy profile. Again, the M-estimator has a bias to larger axion
masses, while the $\langle\sigma_{\rm los}^2\rangle$-Fit recovers anunbiased value. Contours indicate the 1 and 2-$\sigma$ confidence levels. axion mass and soliton density are given in units $10^{-22} {\rm eV}$ and $2.42\times10^{9} M_{\odot} {\rm kpc^{-3}}$
respectively.}
 \end{center}
\end{figure*}

Here we present the triangle plots from the statistical analysis of the mock data presented in Sec.~\ref{sec:analysis-mock-data} generated as isotropic, $\beta=0$, Fig.\ref{fig:triangle1}, and with an Osipkov-Merrit anisotropy profile Fig.\ref{fig:triangle}. In both cases the analysis was done with the two estimators: M-estimator (green line) and $\langle\sigma_{\rm
  los}^2\rangle$-fit (blue lines). We can see that for all the mocks the $\langle\sigma_{\rm
  los}^2\rangle$-fit recovers very well the true values of the axion mass and the central density. These fits included the parameters corresponding to the external NFW part of the profile, however they are not shown as they are fully unconstrained.    
  
\section{ CMB Optical Depth Constraints }
\label{sec:tau}
On of the constraints that can be set by the  epoch of reionization is given by the CMB Thompson scattering optical depth, $\tau$, which is an integral over the reionization history to redshift z,  given by:
\begin{equation}
\tau(z)=\int_{0}{z} \frac{c (1+z')^2}{H(z')} Q_{HII}(z') \sigma_T \bar{n}_{H} (1+\eta Y/4X) dz', 
\end{equation}
where the function $Q_{II}(z)$, volume-filling fraction of ionized hydrogen, and the mean comoving hydrogen number $\bar{n}_H$,  encodes the reionization history. Here, $c$ is the speed of light, $H(z)$ is the Hubble parameter, $\sigma_T$ is the Thompson scattering cross section, and $\eta$ corresponds to the state of the the Helium. The optical depth depend, in general, on the properties of the assumed DM model, mainly through the shape of the linear mass power spectrum. See  \citep{2015MNRAS.450..209B} for a detailed computation of the optical depth in the axion Dark Matter Framework. 

\begin{figure}
\begin{center}
\includegraphics[width=0.5\textwidth]{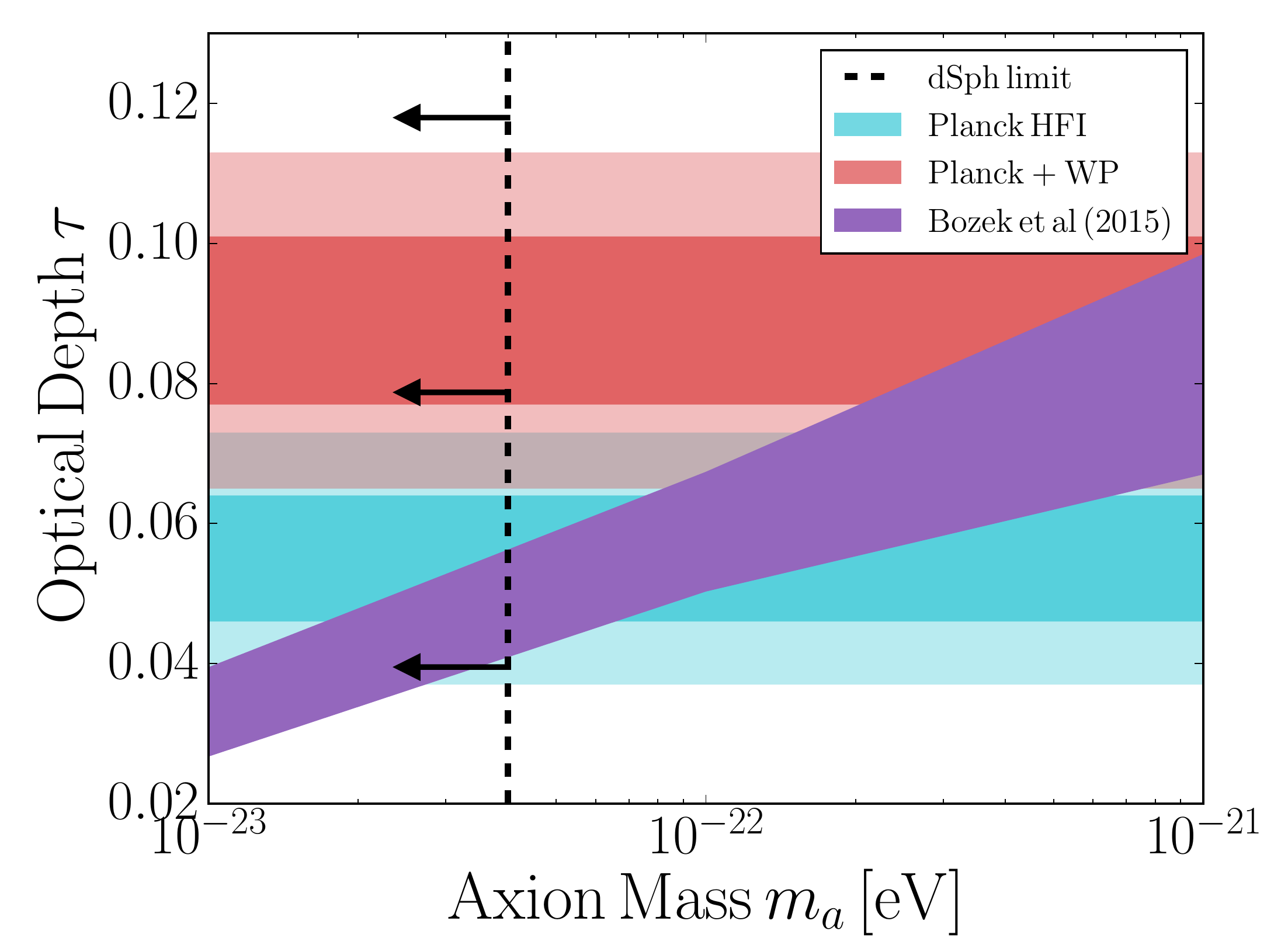}
\end{center}
 \caption{The optical depth to reionization, $\tau$, computed in concordance models for ULAs, where the error band represents systematic modelling uncertainty \citep{2015MNRAS.450..209B}. The 
horizontal bands represent 1 and 2-$\sigma$ constraints on $\tau$ from different CMB polarization power spectrum measurements. The axion mass limit from dSphs produces a reionization history 
consistent with the recent Planck HFI results, but in considerable tension with earlier results from Planck+WP.}
\label{fig:optical_depth}
\end{figure}

Constraints on the CMB Thompson scattering optical depth offers an interesting window onto ULAs in dSphs, as demonstrated in Fig.~\ref{fig:optical_depth}. In this figure we collate the results of \citep{2015MNRAS.450..209B}  for $\tau$ based on concordance reionization models for ULAs, with error bands representing the modeling uncertainty. We compare these results to two different values for $\tau$ determined from CMB temperature and polarization power spectrum measurements. The combination of \emph{Planck} temperature power spectrum and WMAP low-$\ell$ polarization \citep[][``Planck+WP''] {2014A&A...571A..16P,2013ApJS..208...20B}, gives $\tau=0.089\pm 0.012$ \citep[the revision of][to $\tau=0.09\pm 0.13$ does not affect our conclusions]{2015PhRvD..91b3518S}. On the other 
hand, the recent \emph{Planck} low-$\ell$ polarization results using the HFI $\tau$ posterior gives a much lower and tighter value of $\tau=0.055\pm0.009$  \citep[][``Planck HFI'']{2016arXiv160502985P}. The limit on ULA mass from our reanalysis of the dSph data, $m_{22}<0.4$, is in considerable tension with the Planck+WP $\tau$ constraints, yet is consistent with the  Planck HFI constraint. This demonstrates the power that future constraints on the epoch of reionization from CMB polarization will have to probe the nature of DM \citep[e.g.][]{2014JCAP...08..010C}, and the importance of understanding possible low-$\ell$ polarization systematic errors that could be causing a tension between \emph{Planck} and WMAP $\tau$ measurements.

\end{document}